\documentclass[a4paper,12pt]{article}

\usepackage{amsfonts}
\usepackage{amssymb}
\usepackage{amsmath}
\usepackage{mathptmx}
\usepackage{algorithm}
\usepackage{algorithmic}

\usepackage{color}
\usepackage{cancel}
\usepackage{bbm}
\usepackage{pstricks}
\usepackage{psfrag}
\usepackage{mathrsfs}
\usepackage{graphicx}
\def\fileversion{97 patch 1}
\def\filedate{1997/04/28}
\csname PSTplotLoaded\endcsname
\let\PSTplotLoaded 

\ifx\PSTricksLoaded \else
  \def\next{\input pstricks.tex }
  \expandafter\next
\fi

\ifx\MultidoLoaded \else
  \def\next{\input multido.tex }
  \expandafter\next
\fi

\edef\TheAtCode{\the\catcode`\@}
\catcode`\@=11

\begingroup
\catcode`\{=13
\catcode`\}=13
\catcode`\(=13
\catcode`\)=13
\catcode`\,=13
\catcode`\!=1
\catcode`\*=2
\catcode`\ =13
\catcode`\_=13
\catcode`\^^M=13
\gdef\pst@datadelimiters!% Begin def
\catcode`\{=13%
\catcode`\}=13%
\catcode`\(=13%
\catcode`\)=13%
\catcode`\,=13%
\catcode`\ =13%
\catcode`\^^M=13%
\def,##1!%
\ifcat\noexpand,\noexpand##1%
\expandafter##1%
\else\space%
D\space##1%
\fi*%
\let(,\let),\let{,\let},\let ,\let^^M,\let_\@empty*% End def
\endgroup
\begingroup
\catcode`\,=13
\catcode`\_=13
\gdef\savedata@#1[#2]{%
  \xdef\pst@tempg{#2_}%
  \endgroup
  \let#1\pst@tempg
  \global\let\pst@tempg\relax
  \ignorespaces}
\gdef\readdata@{%
  \read1 to \pst@tempa
  \expandafter\readdata@@\pst@tempa_\@nil
  \ifeof1\else\expandafter\readdata@\fi}
\gdef\pst@@readfile#1#2\@nil{\addto@pscode{,#1#2}}%
\gdef\readdata@@#1#2\@nil{\xdef\pst@tempg{\pst@tempg,#1#2}}%
\endgroup

\def\readdata#1#2{%
  \openin1=#2
  \begingroup
  \def\pst@tempg{}%
  \ifeof1
    \@pstrickserr{Data file `#2' not found.}\@ehpa
  \else
    \pst@datadelimiters
    \catcode`\[=1
    \catcode`\]=2
    \readdata@%
  \fi
  \endgroup
  \let#1\pst@tempg
  \global\let\pst@tempg\relax
  \ignorespaces}

\def\pst@readfile#1{{\let\readdata@@\pst@@readfile\readdata\pst@tempg{#1}}}
\def\pst@altreadfile#1{%
  \openin1=#1
  \ifeof1
    \@pstrickserr{Data file `#1' not found.}\@ehpa
  \else
    \catcode`\{=10
    \catcode`\}=10
    \catcode`\(=10
    \catcode`\)=10
    \catcode`\,=10
    \catcode`\^^M=10
    \catcode`\[=1
    \catcode`\]=2
    \pst@@altreadfile
  \fi}
\def\pst@@altreadfile{%
  \read1 to \pst@tempg
  \expandafter\pst@@@altreadfile\pst@tempg\@empty\@nil
  \ifeof1\else\expandafter\pst@@@altreadfile\fi}
\def\pst@@@altreadfile#1#2\@nil{\addto@pscode{#1#2}}%

\def\savedata#1{\begingroup\pst@datadelimiters\savedata@{#1}}

\def\beginplot@line{\begin@OpenObj}
\def\endplot@line{\psline@ii}
\def\beginplot@polygon{\begin@ClosedObj}
\def\endplot@polygon{\pspolygon@ii}
\def\beginplot@curve{\begin@OpenObj}
\def\endplot@curve{\pscurve@ii}
\def\beginplot@ecurve{\begin@OpenObj}
\def\endplot@ecurve{\psecurve@ii}
\def\beginplot@ccurve{\begin@ClosedObj}
\def\endplot@ccurve{\psccurve@ii}
\def\beginplot@dots{\begin@SpecialObj}
\def\endplot@dots{\psdots@ii}
\def\beginplot@bezier{\begin@OpenObj}
\def\endplot@bezier{\psbezier@ii}
\def\beginplot@cbezier{\begin@ClosedObj}
\def\endplot@cbezier{\pscbezier@ii}

\def\psset@plotstyle#1{%
  \@ifundefined{beginplot@#1}%
    {\@pstrickserr{Plot style `#1' not defined}\@eha}%
    {\edef\psplotstyle{#1}}}
\psset@plotstyle{line}

\def\psset@plotpoints#1{%
  \pst@cntg=#1\relax
  \ifnum\pst@cntg<2
    \@pstrickserr{plotpoints parameter must be at least 2}\@ehpa
  \else
    \advance\pst@cntg-1
    \edef\psk@plotpoints{\the\pst@cntg\space}%
  \fi}
\psset@plotpoints{50}

\def\beginqp@line{\pst@oplineto}
\def\doqp@line{L }
\def\endqp@line{\end@OpenObj}%
\def\testqp@line{%
  \ifdim\pslinearc>\z@\else
    \ifshowpoints\else
      \ifx\psk@arrowA\@empty
        \ifx\psk@arrowB\@empty
          \@psttrue
        \fi
      \fi
    \fi
  \fi}

\def\beginqp@polygon{moveto }
\def\doqp@polygon{L }
\def\endqp@polygon{%
  \addto@pscode{closepath}%
  \end@ClosedObj}
\def\testqp@polygon{%
  \ifdim\pslinearc>\z@\else
    \ifshowpoints\else
      \@psttrue
    \fi
  \fi}

\def\beginqp@dots{%
  \psk@dotsize
  \@nameuse{psds@\psk@dotstyle}
  /TheDot { gsave \psk@dotangle \psk@dotscale Dot grestore } def
  TheDot }
\def\doqp@dots{TheDot }
\def\endqp@dots{\end@SpecialObj}
\def\testqp@dots{\@psttrue}

\def\beginqp@bezier{/n 0 def \pst@oplineto}
\def\doqp@bezier{/n n 1 add def n 3 mod 0 eq { curveto } if }
\def\endqp@bezier{%
  \addto@pscode{n 3 mod { pop pop } repeat}
  \end@OpenObj}%
\def\testqp@bezier{%
  \ifshowpoints\else
    \ifx\psk@arrowA\@empty
      \ifx\psk@arrowB\@empty
        \@psttrue
      \fi
    \fi
  \fi}

\def\beginqp@cbezier{/n 0 def moveto }
\def\doqp@cbezier{\doqp@bezier}
\def\endqp@cbezier{%
  \addto@pscode{n 3 mod { pop pop } repeat closepath}
  \end@ClosedObj}%
\def\testqp@cbezier{\ifshowpoints\else\@psttrue\fi}

\def\dataplot{\def\pst@par{}\pst@object{dataplot}}
\def\dataplot@i#1{%
  \pst@killglue
  \begingroup
    \use@par
    \@pstfalse
    \@nameuse{testqp@\psplotstyle}%
    \if@pst
      \dataplot@ii{\addto@pscode{#1}}%
    \else
      \listplot@ii{\addto@pscode{#1}}%
    \fi
  \endgroup
  \ignorespaces}
\def\dataplot@ii#1{%
  \@nameuse{beginplot@\psplotstyle}%
    \addto@pscode{%
      /Dx { \pst@number\psxunit mul /D { Dy } def } def
      /Dy { \pst@number\psyunit mul Do /D { Dx } def } def
      /D { /D { Dx } def } def
      /Do {
        \@nameuse{beginqp@\psplotstyle}%
        /Do { \@nameuse{doqp@\psplotstyle}} def
      } def}%
    #1%
    \addto@pscode{D}%
  \@nameuse{endqp@\psplotstyle}}

\def\fileplot{\def\pst@par{}\pst@object{fileplot}}
\def\fileplot@i#1{%
  \pst@killglue
  \begingroup
    \use@par
    \@pstfalse
    \@nameuse{testqp@\psplotstyle}%
    \if@pst
      \dataplot@ii{\pst@readfile{#1}}%
    \else
      \listplot@ii{\pst@altreadfile{#1}}%
    \fi
  \endgroup
  \ignorespaces}

\pst@def{ScalePoints}<%
  /y ED /x ED
  counttomark dup dup cvi eq not { exch pop } if
  /m exch def /n m 2 div cvi def
  n { y mul m 1 roll x mul m 1 roll /m m 2 sub def } repeat>

\def\listplot{\def\pst@par{}\pst@object{listplot}}
\def\listplot@i#1{\listplot@ii{\addto@pscode{#1}}}
\def\listplot@ii#1{%
  \@nameuse{beginplot@\psplotstyle}%
    \addto@pscode{/D {} def mark}%
    #1%
    \addto@pscode{\pst@number\psxunit \pst@number\psyunit \tx@ScalePoints}%
  \@nameuse{endplot@\psplotstyle}}

\def\psplotinit#1{\xdef\psplot@init{#1 }}
\def\psplot@init{}

\def\psplot{\def\pst@par{}\pst@object{psplot}}
\def\psplot@i#1#2#3{%
  \pst@killglue
  \begingroup
    \use@par
    \@nameuse{beginplot@\psplotstyle}%
    \addto@pscode{%
      \psplot@init
      /x #1 def
      /x1 #2 def
      /dx x1 x sub \psk@plotpoints div def
      /xy {
        x \pst@number\psxunit mul
        #3 \pst@number\psyunit mul
      } def}%
    \gdef\psplot@init{}%
    \@pstfalse
    \@nameuse{testqp@\psplotstyle}%
    \if@pst
      \psplot@ii
    \else
      \psplot@iii
    \fi
  \endgroup
  \ignorespaces}
\def\psplot@ii{%
    \addto@pscode{%
      xy \@nameuse{beginqp@\psplotstyle}
      \psk@plotpoints 1 sub {
        /x x dx add def
        xy \@nameuse{doqp@\psplotstyle}
      } repeat
      /x x1 def
      xy \@nameuse{doqp@\psplotstyle}}%
  \@nameuse{endqp@\psplotstyle}}
\def\psplot@iii{%
    \addto@pscode{%
      mark
      /n 2 def
      \psk@plotpoints {
        xy
        n 2 roll
        /n n 2 add def
        /x x dx add def
      } repeat
      /x x1 def
      xy
      n 2 roll}%
  \@nameuse{endplot@\psplotstyle}}

\def\parametricplot{\def\pst@par{}\pst@object{parametricplot}}
\def\parametricplot@i#1#2#3{%
  \pst@killglue
  \begingroup
    \use@par
    \@nameuse{beginplot@\psplotstyle}%
    \addto@pscode{%
      \psplot@init
      /t #1 def
      /t1 #2 def
      /dt t1 t sub \psk@plotpoints div def
      /xy {
        #3
        \pst@number\psyunit mul exch
        \pst@number\psxunit mul exch
      } def}%
    \gdef\psplot@init{}%
    \@pstfalse
    \@nameuse{testqp@\psplotstyle}%
    \if@pst
      \parametricplot@ii
    \else
      \parametricplot@iii
    \fi
  \endgroup
  \ignorespaces}
\def\parametricplot@ii{%
    \addto@pscode{%
      xy \@nameuse{beginqp@\psplotstyle}
      \psk@plotpoints 1 sub {
        /t t dt add def
        xy \@nameuse{doqp@\psplotstyle}
      } repeat
      /t t1 def
      xy \@nameuse{doqp@\psplotstyle}}%
  \@nameuse{endqp@\psplotstyle}}
\def\parametricplot@iii{%
    \addto@pscode{%
      mark
      /n 2 def
      \psk@plotpoints {
        xy
        n 2 roll
        /n n 2 add def
        /t t dt add def
      } repeat
      /t t1 def
      xy
      n 2 roll}%
  \@nameuse{endplot@\psplotstyle}}

\def\pst@ticks#1#2#3#4{%
  \begin@SpecialObj
    \addto@pscode{%
      #1 rotate
      /n #3 def
      /dx #2 def
      n 0 lt { /dx dx neg def /n n neg def } if
      /y2 \psk@ticksize CLW 2 div add def
      /y1 y2 neg def
      \ifnum\psk@tickstyle=1
        \ifdim#4<\z@ /y2 \else /y1 \fi 0 def
      \else
        \ifnum\psk@tickstyle=-1
          \ifdim#4<\z@ /y1 \else /y2 \fi 0 def
        \fi
      \fi
      /x dx def
      n { x y1 moveto x y2 lineto stroke /x x dx add def } repeat}%
  \end@SpecialObj}

\def\psset@ticksize#1{\pst@getlength{#1}\psk@ticksize}
\psset@ticksize{3pt}

\def\psset@tickstyle#1{\pst@expandafter\psset@@tickstyle{#1}\@nil}
\def\psset@@tickstyle#1#2\@nil{%
  \ifx#1f\let\psk@tickstyle\z@\else
    \ifx#1t\let\psk@tickstyle\@ne\else
      \ifx#1b\let\psk@tickstyle\m@ne\else
        \@pstrickserr{Bad tick style: `#1#2'}\@ehpa
  \fi\fi\fi}
\psset@tickstyle{full}

\def\psset@ticks#1{\pst@expandafter\psset@@ticks{#1}\@nil\psk@ticks}
\def\psset@@ticks#1#2\@nil#3{%
  \ifx#1a\let#3\z@\else
    \ifx#1x\let#3\@ne\else
      \ifx#1y\let#3\tw@\else
        \ifx#1n\let#3\thr@@\else
          \@pstrickserr{Bad argument: `#1#2'}\@ehpa
  \fi\fi\fi\fi}
\psset@ticks{all}

\def\psset@labels#1{\pst@expandafter\psset@@ticks{#1}\@nil\psk@labels}
\psset@labels{all}

\def\psset@Ox#1{\edef\psk@Ox{#1}}
\psset@Ox{0}
\def\psset@Dx#1{\edef\psk@Dx{#1}}
\psset@Dx{1}
\def\psset@dx#1{%
  \pssetxlength\pst@dimg{#1}%
  \edef\psk@dx{\number\pst@dimg}}
\psset@dx{0}

\def\psset@Oy#1{\edef\psk@Oy{#1}}
\psset@Oy{0}
\def\psset@Dy#1{\edef\psk@Dy{#1}}
\psset@Dy{1}
\def\psset@dy#1{%
  \pssetylength\pst@dimg{#1}%
  \edef\psk@dy{\number\pst@dimg}}
\psset@dy{0}

\newif\ifshoworigin
\def\psset@showorigin#1{\@nameuse{showorigin#1}}
\psset@showorigin{true}

\def\psaxes{\def\pst@par{}\pst@object{psaxes}}
\def\psaxes@i{\pst@getarrows\psaxes@ii}
\def\psaxes@ii(#1){\@ifnextchar({\psaxes@iii(#1)}{\psaxes@iv(0,0)(0,0)(#1)}}
\def\psaxes@iii(#1)(#2){%
  \@ifnextchar(%
    {\psaxes@iv(#1)(#2)}%
    {\psaxes@iv(#1)(#1)(#2)}}
\def\psaxes@iv(#1,#2)(#3,#4)(#5,#6){%
  \setbox\pst@hbox=\hbox\bgroup
    \use@par
    \pssetxlength\pst@dimg{#1}% o-x
    \pssetylength\pst@dimh{#2}% o-y
    \pssetxlength\pst@dima{#3}% bl-x
    \pssetylength\pst@dimb{#4}% bl-y
    \pssetxlength\pst@dimc{#5}% ur-x
    \pssetylength\pst@dimd{#6}% ur-y
    \advance\pst@dima-\pst@dimg % Dist. from bl-x to o-x
    \advance\pst@dimb-\pst@dimh % Dist. from bl-y to o-y
    \advance\pst@dimc-\pst@dimg % Dist. from ur-x to o-x
    \advance\pst@dimd-\pst@dimh % Dist. from ur-y to o-y
    \@nameuse{psxs@\psk@axesstyle}%
    \advance\pslabelsep.5\pslinewidth
    \begingroup
      \ifdim\pst@dimb=\z@\else\showoriginfalse\fi
      \ifnum\psk@dx=\z@
        \pst@dimg=\psk@Dx\psxunit
        \edef\psk@dx{\number\pst@dimg}%
      \fi
      \ifnum\psk@ticks<\tw@
        \ifnum\psk@tickstyle>\z@\else
          \advance\pslabelsep\psk@ticksize\p@
        \fi
      \fi
      \pst@hlabels\pst@dimc\psk@arrowB
      \pst@hlabels\pst@dima\psk@arrowA
    \endgroup
    \begingroup
      \ifdim\pst@dima=\z@\else\showoriginfalse\fi
      \ifnum\psk@dy=\z@
         \pst@dimg=\psk@Dy\psyunit
         \edef\psk@dy{\number\pst@dimg}%
      \fi
      \ifodd\psk@ticks\else
        \ifnum\psk@tickstyle>\z@\else
          \advance\pslabelsep\psk@ticksize\p@
        \fi
      \fi
      \pst@vlabels\pst@dimd\psk@arrowB
      \pst@vlabels\pst@dimb\psk@arrowA
    \endgroup
  \egroup
  \pssetxlength\pst@dimg{#1}%
  \pssetylength\pst@dimh{#2}%
  \leavevmode\psput@cartesian\pst@hbox
  \ignorespaces}

\def\psxs@axes{%
  \psxs@@axes\pst@dima\pst@dimc{}%
  \psxs@@axes\pst@dimb\pst@dimd{exch}}
\def\psxs@@axes#1#2#3{%
  \begin@SpecialObj
    \ifdim#1=\z@
      \def\psk@arrowA{C}%
    \else
      \ifdim#2=\z@
        \def\psk@arrowB{C}%
      \fi
    \fi
    \let\pst@linetype\pst@arrowtype
    \pst@addarrowdef
    \addto@pscode{%
      \pst@number#2 0 #3
      \pst@number#1 0 #3
      ArrowA
      CP 4 2 roll
      ArrowB
      L
      pop pop}%
    \pst@stroke
  \end@SpecialObj}

\def\psxs@frame{%
  \begin@SpecialObj
    \addto@pscode{%
      0 0 moveto \pst@number\pst@dimc 0 L
      0 \pst@number\pst@dimd 2 copy rlineto L closepath}%
    \pst@stroke
    \psk@fillstyle
  \end@SpecialObj
  \let\psk@arrowA\@empty
  \let\psk@arrowB\@empty}

\def\psset@axesstyle#1{%
  \@ifundefined{psxs@#1}%
    {\@pstrickserr{Axes style `#1' not defined}\@eha}%
    {\edef\psk@axesstyle{#1}}}
\psset@axesstyle{axes}

\def\psxs@none{\let\psk@arrowA\@empty\let\psk@arrowB\@empty}

\def\pst@hlabels#1#2{%
  \ifdim#1=\z@\else
    \ifx#2\empty\else
      \advance#1\ifdim#1>\z@-\fi7\pslinewidth
    \fi
    \pst@cnta=#1\relax                % Distance (in sp) to end.
    \divide\pst@cnta\psk@dx\relax     % Number of ticks/labels
    \ifnum\pst@cnta=\z@\else
      \pst@dimb=\psk@dx sp            % Space between ticks.
      \ifnum\psk@ticks<\tw@
        \pst@ticks{0}{\pst@number\pst@dimb}{\the\pst@cnta}{\pst@dimd}%
      \fi
      \ifnum\psk@labels<\tw@ \pst@@hlabels\fi
      \showoriginfalse
    \fi
  \fi}

\def\pst@@hlabels{%
  \vbox to\z@{%
    \ifdim\pst@dimd>\z@\vskip\pslabelsep\else\vss\fi
    \ifnum\pst@cnta<\z@
      \pst@dimb=-\pst@dimb
    \fi
    \hbox to\z@{%
      \ifshoworigin\hbox to \z@{\hss\pshlabel{\psk@Ox}\hss}\fi
      \mmultido
        {\n=\psk@Ox+\psk@Dx}%
        {\pst@cnta}%
        {\hskip\pst@dimb\hbox to \z@{\hss\pshlabel{\n}\hss}}%
      \hss}%
    \ifdim\pst@dimd>\z@\vss\else\vskip\pslabelsep\fi}}%

\def\pshlabel#1{$#1$}

\def\pst@vlabels#1#2{%
  \ifdim#1=\z@\else
    \ifx#2\empty\else
      \advance#1\ifdim#1>\z@-\fi7\pslinewidth
    \fi
    \pst@cnta=#1\relax                % Distance (in sp) to end.
    \divide\pst@cnta\psk@dy\relax     % Number of ticks/labels
    \ifnum\pst@cnta=\z@\else
      \pst@dima=\psk@dy sp            % Space between ticks.
      \ifodd\psk@ticks\else
        \pst@ticks{90}{\pst@number\pst@dima}{\the\pst@cnta}{-\pst@dimc}%
      \fi
      \ifodd\psk@labels\else\pst@@vlabels\fi
      \showoriginfalse
    \fi
  \fi}

\def\pst@@vlabels{%
  \vbox to\z@{%
    \ifnum\pst@cnta>\z@
      \pst@dima=-\pst@dima
    \fi
    \offinterlineskip
    \ifshoworigin
      \vbox to \z@{\vss\hbox to\z@{%
        \ifdim\pst@dimc>\z@\hss\else\hskip\pslabelsep\fi
        \psvlabel{\psk@Oy}%
        \ifdim\pst@dimc>\z@\hskip\pslabelsep\else\hss\fi}\vss}%
    \fi
    \mmultido
      {\n=\psk@Oy+\psk@Dy}%
      {\pst@cnta}%
      {\vbox to\pst@dima{\vss}\vbox to \z@{\vss\hbox to\z@{%
        \ifdim\pst@dimc>\z@\hss\else\hskip\pslabelsep\fi
        \psvlabel{\n}%
        \ifdim\pst@dimc>\z@\hskip\pslabelsep\else\hss\fi}\vss}}%
    \vss}}

\def\psvlabel#1{$#1$}

\catcode`\@=\TheAtCode\relax
 
\def\fileversion{97 patch 6}
\def\filedate{1998/04/28}
\csname PSTnodesLoaded\endcsname
\let\PSTnodesLoaded 
\ifx\PSTricksLoaded \else\input pstricks.tex\fi\relax
\edef\TheAtCode{\the\catcode`\@}
\catcode`\@=11
\pstheader{pst-node.pro}
\def\pst@nodedict{tx@NodeDict begin }
\def\pst@zapspace#1 #2{%
#1%
\ifx#2\@empty\else\expandafter\pst@zapspace\fi
#2}
\def\pst@getnode#1#2{%
\pst@expandafter\pst@@getnode{#1},,\@nil#2}
\def\pst@@getnode#1,#2,#3\@nil#4{%
\ifx\@empty#3\@empty
\edef#4{/N@\pst@zapspace#1 \@empty\space}%
\else
\pst@cntg=#1\relax
\pst@cnth=#2\relax
\edef#4{/N@M-\ifnum\psmatrixcnt=\z@ 1\else\the\psmatrixcnt\fi
-\the\pst@cntg-\the\pst@cnth\space}%
\fi}
\def\tx@NewNode{NewNode }
\def\pst@newnode#1#2#3#4{%
\leavevmode
\pst@getnode{#1}\pst@thenode
\pst@Verb{%
\pst@nodedict
{#3}
\ifx\psk@name\relax false \else \psk@name true \fi
\pst@thenode
#2
{#4}
\tx@NewNode
end}%
\global\let\psk@name\relax
\pstree@nodehook
\global\let\pstree@nodehook\relax}
\let\pstree@nodehook\relax
\newif\ifnodealign
\def\psset@nodealign#1{\@nameuse{nodealign#1}}
\psset@nodealign{false}
\def\pst@nodealign{%
\pst@dimg=\ht\pst@hbox
\advance\pst@dimg-\dp\pst@hbox
\divide\pst@dimg2
\lower\pst@dimg}
\def\tx@InitPnode{InitPnode }
\def\pnode{\@ifnextchar({\pnode@}{\pnode@(0,0)}}
\def\pnode@(#1)#2{%
\pst@@getcoor{#1}%
\pst@newnode{#2}{10}{\pst@coor}{\tx@InitPnode}%
\ignorespaces}
\def\tx@InitCnode{InitCnode }
\def\cnode{\pst@object{cnode}}
\def\cnode@i{\@ifnextchar({\cnode@ii}{\cnode@ii(0,0)}}
\def\cnode@ii(#1)#2#3{%
\leavevmode
\hbox{%
\use@par
\pst@@getcoor{#1}%
\pssetlength\pst@dimc{#2}%
\pst@dimg=\psk@dimen\pslinewidth
\advance\pst@dimc-\pst@dimg
\advance\pst@dimc.5\pslinewidth
\ifnodealign
\kern\pst@dimc
\vrule width\z@ height \pst@dimc depth \pst@dimc
\fi
\pscircle@do(#1){#2}%
\pst@newnode{#3}{11}{\pst@coor \pst@number\pst@dimc}{\tx@InitCnode}%
\ifnodealign\kern\pst@dimc\fi}%
\ignorespaces}
\def\Cnode{\pst@object{Cnode}}
\def\Cnode@i{\@ifnextchar({\Cnode@ii}{\Cnode@ii(0,0)}}
\def\Cnode@ii(#1)#2{\cnode@ii(#1){\psk@radius}{#2}}%
\def\cnodeput{\pst@object{cnodeput}}
\def\cnodeput@i{\@ifnextchar({\cnodeput@iii}{\cnodeput@ii}}
\def\cnodeput@ii#1{%
\addto@par{rot={#1}}%
\@ifnextchar({\cnodeput@iii}{\cnodeput@iii(\z@,\z@)}}
\def\cnodeput@iii(#1)#2{%
\pst@killglue
\@fixedradiusfalse
\def\pst@nodehook{\cnodeput@iv{#2}}%
\pst@makebox{\cput@v{#1}}}
\def\cnodeput@iv#1{%
\pst@newnode{#1}{11}{\pscirclebox@iv \pst@number\pslinewidth add}%
{\tx@InitCnode}%
\global\let\pst@nodehook\relax}
\def\Cnodeput{\pst@object{Cnodeput}}
\def\Cnodeput@i{\@ifnextchar({\Cnodeput@iii}{\Cnodeput@ii}}
\def\Cnodeput@ii#1{%
\addto@par{rot={#1}}%
\@ifnextchar({\Cnodeput@iii}{\Cnodeput@iii(\z@,\z@)}}
\def\Cnodeput@iii(#1)#2{%
\pst@killglue
\@fixedradiustrue
\def\pst@nodehook{\Cnodeput@iv{#2}}%
\pst@makebox{\cput@iv{#1}}}
\def\Cnodeput@iv#1{%
\pst@newnode{#1}{11}{%
\pst@number{\wd\pst@hbox} 2 div \pst@number\pst@dima % x y
\pst@number\pst@dimb \pst@number\pslinewidth \psk@dimen .5 sub mul sub }% r
{\tx@InitCnode}%
\global\let\pst@nodehook\relax}
\def\circlenode{\pst@object{circlenode}}
\def\circlenode@i#1{\pst@makebox{\circlenode@ii{#1}}}
\def\circlenode@ii#1{%
\begingroup
\pst@useboxpar
\setbox\pst@hbox=\hbox{%
\cnodeput@iv{#1}%
\pscirclebox@iii
\box\pst@hbox}%
\ifnodealign \psboxseptrue \fi
\ifpsboxsep \pscirclebox@sep \fi
\leavevmode
\ifnodealign\pst@nodealign\fi
\box\pst@hbox
\endgroup}
\def\Circlenode{\pst@object{Circlenode}}
\def\Circlenode@i#1{\pst@makebox{\Circlenode@ii{#1}}}
\def\Circlenode@ii#1{%
\begingroup
\pst@useboxpar
\pst@dima=\ht\pst@hbox
\advance\pst@dima\dp\pst@hbox
\divide\pst@dima\tw@
\pssetlength\pst@dimb\psk@radius
\setbox\pst@hbox=\hbox{%
\Cnodeput@iv{#1}%
\pscircle(.5\wd\pst@hbox,\pst@dima){\pst@dimb}%
\box\pst@hbox}%
\ifnodealign \psboxseptrue \fi
\ifpsboxsep \psCirclebox@sep \fi
\leavevmode
\ifnodealign\pst@nodealign\fi
\box\pst@hbox
\endgroup}
\def\tx@GetRnodePos{GetRnodePos }
\def\tx@InitRnode{InitRnode }
\def\rnode{\@ifnextchar[{\rnode@i}{\def\pst@par{}\rnode@ii}}
\def\rnode@i[#1]{\def\pst@par{ref=#1}\rnode@ii}
\def\rnode@ii#1{\pst@makebox{\rnode@iii\rnode@iv{#1}}}
\def\rnode@iii#1#2{%
\leavevmode
\begingroup
\pst@useboxpar
#1%
\if@star\pst@starbox\fi
\ifnodealign\lower\pst@dimb\fi
\hbox{%
\pst@newnode{#2}{16}{%
\pst@number{\ht\pst@hbox}%
\pst@number{\dp\pst@hbox}%
\pst@number{\wd\pst@hbox}%
\pst@number\pst@dima
\pst@number\pst@dimb}%
{\tx@InitRnode}%
\box\pst@hbox}%
\endgroup}
\def\rnode@iv{%
\pst@dima=\psk@xref\wd\pst@hbox
\ifx\psk@yref\relax
\pst@dimb=\z@
\else
\pst@dimb=\ht\pst@hbox
\advance\pst@dimb\dp\pst@hbox
\pst@dimb=\psk@yref\pst@dimb
\advance\pst@dimb-\dp\pst@hbox
\fi}
\def\psset@href#1{\pst@checknum{#1}\psk@href}
\psset@href{0}
\def\psset@vref#1{\def\psk@vref{#1}}
\psset@vref{.7ex}
\def\Rnode{\pst@object{Rnode}}
\def\Rnode@i#1{\pst@makebox{\rnode@iii\Rnode@ii{#1}}}
\def\Rnode@ii{%
\use@par
\pst@dima=\psk@href\wd\pst@hbox
\advance\pst@dima\wd\pst@hbox
\divide\pst@dima 2
\pssetlength\pst@dimb{\psk@vref}}
\def\tx@DiaNodePos{DiaNodePos }
\def\dianode{\pst@object{dianode}}
\def\dianode@i#1{\pst@makebox{\dianode@ii{#1}}}
\def\dianode@ii#1{%
\begingroup
\pst@useboxpar
\psdiabox@iii
\setbox\pst@hbox=\hbox{%
\pst@newnode{#1}{14}{}{%
/X \pst@number\pst@dima def
/Y \pst@number\pst@dimb def
/w \pst@number\pst@dimc 2 mul def
/h \pst@number\pst@dimd 2 mul def
/NodePos { \tx@DiaNodePos } def}%
\box\pst@hbox}%
\ifnodealign\psboxseptrue\fi
\ifpsboxsep\psdiabox@sep\fi
\leavevmode
\ifnodealign\lower\pst@dimb\fi
\box\pst@hbox
\endgroup}
\def\tx@TriNodePos{TriNodePos }
\def\tx@InitTriNode{InitTriNode }
\def\trinode{\pst@object{trinode}}
\def\trinode@i#1{\pst@makebox{\trinode@ii{#1}}}
\def\trinode@ii#1{%
\begingroup
\pst@useboxpar
\pstribox@iii
\setbox\pst@hbox=\hbox{%
\pst@newnode{#1}{14}{}{%
\pst@number\pst@dimc
\pst@number\pst@dimd
\ifodd\psk@trimode
exch
\pst@number\pst@dima
\else
\pst@number\pst@dimb
\fi
\psk@trimode
\pst@number{\wd\pst@hbox}%
\pst@number{\ht\pst@hbox}%
\pst@number{\dp\pst@hbox}%
\tx@InitTriNode}%
\box\pst@hbox}%
\ifnodealign\psboxseptrue\fi
\ifpsboxsep\pstribox@sep\fi
\leavevmode
\ifnodealign\lower\pst@tempa\fi
\box\pst@hbox
\endgroup}
\def\tx@OvalNodePos{OvalNodePos }
\def\ovalnode{\pst@object{ovalnode}}
\def\ovalnode@i#1{\pst@makebox{\ovalnode@ii{#1}}}
\def\ovalnode@ii#1{%
\begingroup
\pst@useboxpar
\psovalbox@iii
\setbox\pst@hbox=\hbox{%
\pst@newnode{#1}{14}{}{%
/X \pst@number\pst@dima def
/Y \pst@number\pst@dimb def
/w \pst@number\pst@dimc def
/h \pst@number\pst@dimd def
/NodePos { \tx@OvalNodePos } def}%
\unhbox\pst@hbox}%
\ifnodealign\psboxseptrue\fi
\ifpsboxsep\psovalbox@sep\fi
\leavevmode
\ifnodealign\lower\pst@dimb\fi
\box\pst@hbox
\endgroup}
\def\dotnode{\pst@object{dotnode}}
\def\dotnode@i{\@ifnextchar({\dotnode@ii}{\dotnode@ii(\z@,\z@)}}
\def\dotnode@ii(#1)#2{%
\leavevmode
\hbox{%
\use@par
\pst@@getcoor{#1}%
\pst@getdotsize
\pstree@nodehook
\ifnodealign
\pst@dima=\pst@dimg
\kern\pst@dima
\vrule width\z@ height \pst@dimh depth \pst@dimh
\fi
\pst@newnode{#2}{14}{}{%
\pst@coor
/Y ED /X ED
/w \pst@number\pst@dimg def
/h \pst@number\pst@dimh def
/NodePos { \tx@OvalNodePos } def}%
\psdot@ii(#1)%
\ifnodealign\kern\pst@dima\fi}%
\ignorespaces}
\def\psset@framesize#1{\pst@expandafter\psset@@framesize{#1} \@nil}
\def\psset@@framesize#1 #2\@nil{%
\pssetlength\pst@dimg{#1}%
\divide\pst@dimg2
\edef\psk@framewidth{\pst@number\pst@dimg}%
\ifx\@empty#2\@empty
\let\psk@frameheight\psk@framewidth
\else
\pssetlength\pst@dimg{#2}%
\divide\pst@dimg2
\edef\psk@frameheight{\pst@number\pst@dimg}%
\fi}
\psset@framesize{10pt}
\def\fnode{\pst@object{fnode}}
\def\fnode@i{\@ifnextchar({\fnode@ii}{\fnode@ii(\z@,\z@)}}
\def\fnode@ii(#1)#2{%
\leavevmode
\pst@killglue
\hbox{%
\use@par
\begin@ClosedObj
\ifnodealign
\kern\psk@framewidth\p@
\vrule width\z@ height \psk@frameheight\p@ depth \psk@frameheight\p@
\edef\pst@coor{0 0 }%
\else
\pst@@getcoor{#1}%
\fi
\pst@newnode{#2}{14}{}{%
\pst@coor
/Y ED /X ED
/d \psk@dimen .5 sub CLW mul neg def
/r \psk@framewidth d add def
/l r neg def
/u \psk@frameheight d add def
/d u neg def
/NodePos { \tx@GetRnodePos } def}%
\addto@pscode{%
/x2 \psk@framewidth CLW \psk@dimen mul sub def
/y2 \psk@frameheight CLW \psk@dimen mul sub def
\pst@coor 2 copy
y2 sub /y1 ED
x2 sub /x1 ED
y2 add /y2 ED
x2 add /x2 ED
\psk@cornersize
1 index 0 eq { pop pop \tx@Rect } { \tx@OvalFrame } ifelse}%
\def\pst@linetype{2}%
\showpointsfalse
\end@ClosedObj
\ifnodealign\kern\psk@framewidth\p@\fi}%
\ignorespaces}
\def\psset@nodesepA#1{%
\pst@getlength{#1}\psk@nodesepA
\def\psk@nodeseptypeA{0 }}
\def\psset@nodesepB#1{%
\pst@getlength{#1}\psk@nodesepB
\def\psk@nodeseptypeB{0 }}
\def\psset@nodesep#1{%
\pst@getlength{#1}\psk@nodesepA
\let\psk@nodesepB\psk@nodesepA
\def\psk@nodeseptypeA{0 }%
\def\psk@nodeseptypeB{0 }}
\psset@nodesep{0pt}
\def\psset@XnodesepA#1{%
\pst@getlength{#1}\psk@nodesepA
\def\psk@nodeseptypeA{1 }}
\def\psset@XnodesepB#1{%
\pst@getlength{#1}\psk@nodesepB
\def\psk@nodeseptypeB{1 }}
\def\psset@Xnodesep#1{%
\pst@getlength{#1}\psk@nodesepA
\let\psk@nodesepB\psk@nodesepA
\def\psk@nodeseptypeA{1 }%
\def\psk@nodeseptypeB{1 }}
\def\psset@YnodesepA#1{%
\pst@getlength{#1}\psk@nodesepA
\def\psk@nodeseptypeA{2 }}
\def\psset@YnodesepB#1{%
\pst@getlength{#1}\psk@nodesepB
\def\psk@nodeseptypeB{2 }}
\def\psset@Ynodesep#1{%
\pst@getlength{#1}\psk@nodesepA
\let\psk@nodesepB\psk@nodesepA
\def\psk@nodeseptypeA{2 }%
\def\psk@nodeseptypeB{2 }}
\def\psset@armA#1{%
\pst@getlength{#1}\psk@armA
\def\psk@armtypeA{0 }}
\def\psset@armB#1{%
\pst@getlength{#1}\psk@armB
\def\psk@armtypeB{0 }}
\def\psset@arm#1{%
\pst@getlength{#1}\psk@armA
\let\psk@armB\psk@armA
\def\psk@armtypeA{0 }%
\def\psk@armtypeB{0 }}
\psset@arm{10pt}
\def\psset@XarmA#1{%
\pst@getlength{#1}\psk@armA
\def\psk@armtypeA{1 }}
\def\psset@XarmB#1{%
\pst@getlength{#1}\psk@armB
\def\psk@armtypeB{1 }}
\def\psset@Xarm#1{%
\pst@getlength{#1}\psk@armA
\let\psk@armB\psk@armA
\def\psk@armtypeA{1 }%
\def\psk@armtypeB{1 }}
\def\psset@YarmA#1{%
\pst@getlength{#1}\psk@armA
\def\psk@armtypeA{2 }}
\def\psset@YarmB#1{%
\pst@getlength{#1}\psk@armB
\def\psk@armtypeB{2 }}
\def\psset@Yarm#1{%
\pst@getlength{#1}\psk@armA
\let\psk@armB\psk@armA
\def\psk@armtypeA{2 }%
\def\psk@armtypeB{2 }}
\def\psset@offsetA#1{\pst@getlength{#1}\psk@offsetA}
\def\psset@offsetB#1{\pst@getlength{#1}\psk@offsetB}
\def\psset@offset#1{\psset@offsetA{#1}\let\psk@offsetB\psk@offsetA}
\psset@offset{0pt}
\def\psset@angleA#1{\pst@getangle{#1}\psk@angleA}
\def\psset@angleB#1{\pst@getangle{#1}\psk@angleB}%
\def\psset@angle#1{%
\pst@getangle{#1}\psk@angleA
\let\psk@angleB\psk@angleA}
\psset@angle{0}
\def\psset@arcangleA#1{\pst@getangle{#1}\psk@arcangleA}
\def\psset@arcangleB#1{\pst@getangle{#1}\psk@arcangleB}%
\def\psset@arcangle#1{%
\pst@getangle{#1}\psk@arcangleA
\let\psk@arcangleB\psk@arcangleA}
\psset@arcangle{8}
\def\psset@ncurvA#1{\pst@checknum{#1}\psk@ncurvA}
\def\psset@ncurvB#1{\pst@checknum{#1}\psk@ncurvB}%
\def\psset@ncurv#1{\psset@ncurvA{#1}\let\psk@ncurvB\psk@ncurvA}
\psset@ncurv{.67}
\def\tx@GetCenter{GetCenter }
\def\tx@XYPos{XYPos }
\def\tx@GetEdge{GetEdge }
\def\tx@AddOffset{AddOffset }
\def\tx@GetEdgeA{GetEdgeA }
\def\tx@GetEdgeB{GetEdgeB }
\def\tx@GetArmA{GetArmA }
\def\tx@GetArmB{GetArmB }
\def\check@arrow#1#2{%
\check@@arrow#2-\@nil
\if@pst
\addto@par{arrows=#2}%
\def\next{#1}%
\else
\def\next{#1{#2}}%
\fi
\next}
\def\check@@arrow#1-#2\@nil{%
\ifx\@nil#2\@nil\@pstfalse\else\@psttrue\fi}
\def\tx@InitNC{InitNC }
\def\nc@object#1#2#3#4#5{%
\csname begin@#1Obj\endcsname
\showpointsfalse
\pst@getnode{#2}\pst@tempa
\pst@getnode{#3}\pst@tempb
\gdef\npos@default{#4 }%
\addto@pscode{%
/NCLW CLW def
\pst@nodedict
\psk@offsetA
\psk@offsetB neg
\psk@nodesepA
\psk@nodesepB
\psk@nodeseptypeA
\psk@nodeseptypeB
\pst@tempa
\pst@tempb
\tx@InitNC { #5 } if
end}%
\def\use@pscode{%
\pst@Verb{gsave \tx@STV newpath \pst@code\space grestore}%
\gdef\pst@code{}}%
\csname end@#1Obj\endcsname
\pst@shortput}
\def\npos@default{.5 }
\def\pc@object#1{%
\@ifnextchar({\pc@@object#1}{\pst@getarrows{\pc@@object#1}}}
\def\pc@@object#1(#2)(#3){%
\pnode(#2){@@A}\pnode(#3){@@B}%
#1{@@A}{@@B}}
\def\tx@LPutLine{LPutLine }
\def\tx@LPutLines{LPutLines }
\def\tx@BezierMidpoint{BezierMidpoint }
\def\tx@HPosBegin{HPosBegin }
\def\tx@HPosEnd{HPosEnd }
\def\tx@HPutLine{HPutLine }
\def\tx@HPutLines{HPutLines }
\def\tx@VPosBegin{VPosBegin }
\def\tx@VPosEnd{VPosEnd }
\def\tx@VPutLine{VPutLine }
\def\tx@VPutLines{VPutLines }
\def\tx@HPutCurve{HPutCurve }
\def\tx@NCCoor{NCCoor }
\def\tx@NCLine{NCLine }
\def\ncline{\pst@object{ncline}}
\def\ncline@i{\check@arrow{\ncline@ii}}
\def\ncline@ii#1#2{\nc@object{Open}{#1}{#2}{.5}{\tx@NCLine}}
\def\pcline{\pst@object{pcline}}
\def\pcline@i{\pc@object\ncline@ii}
\def\ncLine{\pst@object{ncLine}}
\def\ncLine@i{\check@arrow{\ncLine@ii}}
\def\ncLine@ii#1#2{\nc@object{Open}{#1}{#2}{.5}%
{\tx@NCLine /LPutPos { xB xA yB yA \tx@LPutLine } def}}
\def\tx@NCLines{NCLines }
\def\nclines{\pst@object{nclines}}
\def\nclines@i{\check@arrow\nclines@ii}
\def\nclines@ii#1#2{%
\begingroup
\use@par
\def\pst@aftercoors{\nclines@iii{#1}{#2}}%
\def\pst@coors{}%
\pst@@getcoors}
\def\nclines@iii#1#2{%
\nc@object{Open}{#1}{#2}{.5}{%
tx@Dict begin \psline@iii pop end
mark \pst@coors \tx@NCLines}%
\endgroup
\ignorespaces}
\def\tx@NCCurve{NCCurve }
\def\nccurve{\pst@object{nccurve}}
\def\nccurve@i{\check@arrow{\nccurve@ii}}
\def\nccurve@ii#1#2{\nc@object{Open}{#1}{#2}{.5}{%
/AngleA \psk@angleA\space def /AngleB \psk@angleB\space def
\psk@ncurvB\space \psk@ncurvA\space
\tx@NCCurve}}
\def\pccurve{\pst@object{pccurve}}
\def\pccurve@i{\pc@object\nccurve@ii}
\def\ncarc{\pst@object{ncarc}}
\def\ncarc@i{\check@arrow{\ncarc@ii}}
\def\ncarc@ii#1#2{\nc@object{Open}{#1}{#2}{.5}{%
yB yA sub xB xA sub \tx@Atan dup
\psk@arcangleA\space add /AngleA ED
\psk@arcangleB\space sub 180 add /AngleB ED
\psk@ncurvB\space \psk@ncurvA\space
\tx@NCCurve}}
\def\pcarc{\pst@object{pcarc}}
\def\pcarc@i{\pc@object\ncarc@ii}
\def\tx@NCAngles{NCAngles }
\def\ncangles{\pst@object{ncangles}}
\def\ncangles@i{\check@arrow{\ncangles@ii}}
\def\ncangles@ii#1#2{%
\nc@object{Open}{#1}{#2}{1.5}{\ncangles@iii \tx@NCAngles}}
\def\ncangles@iii{%
tx@Dict begin \psline@iii pop end
/AngleA \psk@angleA def
/AngleB \psk@angleB def
/ArmA \psk@armA def
/ArmB \psk@armB def
/ArmTypeA \psk@armtypeA def
/ArmTypeB \psk@armtypeB def }
\def\pcangles{\pst@object{pcangles}}
\def\pcangles@i{\pc@object\ncangles@ii}
\def\tx@NCAngle{NCAngle }
\def\ncangle{\pst@object{ncangle}}
\def\ncangle@i{\check@arrow{\ncangle@ii}}
\def\ncangle@ii#1#2{%
\nc@object{Open}{#1}{#2}{1.5}{\ncangles@iii \tx@NCAngle}}
\def\pcangle{\pst@object{pcangle}}
\def\pcangle@i{\pc@object\ncangle@ii}
\def\tx@NCBar{NCBar }
\def\ncbar{\pst@object{ncbar}}
\def\ncbar@i{\check@arrow{\ncbar@ii}}
\def\ncbar@ii#1#2{\nc@object{Open}{#1}{#2}{1.5}{%
\ncangles@iii /AngleB \psk@angleA def \tx@NCBar}}
\def\pcbar{\pst@object{pcbar}}
\def\pcbar@i{\pc@object\ncbar@ii}
\def\tx@NCDiag{NCDiag }
\def\ncdiag{\pst@object{ncdiag}}
\def\ncdiag@i{\check@arrow{\ncdiag@ii}}
\def\ncdiag@ii#1#2{%
\nc@object{Open}{#1}{#2}{1.5}{\ncangles@iii \tx@NCDiag}}
\def\pcdiag{\pst@object{pcdiag}}
\def\pcdiag@i{\pc@object\ncdiag@ii}
\def\tx@NCDiagg{NCDiagg }
\def\ncdiagg{\pst@object{ncdiagg}}
\def\ncdiagg@i{\check@arrow{\ncdiagg@ii}}
\def\ncdiagg@ii#1#2{%
\nc@object{Open}{#1}{#2}{.5}{\ncangles@iii \tx@NCDiagg}}
\def\pcdiagg{\pst@object{pcdiagg}}
\def\pcdiagg@i{\pc@object\ncdiagg@ii}
\def\tx@NCLoop{NCLoop }
\def\psset@loopsize#1{\pst@getlength{#1}\psk@loopsize}
\psset@loopsize{1cm}
\def\ncloop{\pst@object{ncloop}}
\def\ncloop@i{\check@arrow{\ncloop@ii}}
\def\ncloop@ii#1#2{%
\nc@object{Open}{#1}{#2}{2.5}%
{\ncangles@iii /loopsize \psk@loopsize def \tx@NCLoop}}
\def\pcloop{\pst@object{pcloop}}
\def\pcloop@i{\pc@object\ncloop@ii}
\def\tx@NCCircle{NCCircle }
\def\nccircle{\pst@object{nccircle}}
\def\nccircle@i{\check@arrow{\nccircle@ii}}
\def\nccircle@ii#1#2{%
\pssetlength\pst@dima{#2}%
\nc@object{Open}{#1}{#1}{.5}{%
/AngleA \psk@angleA def
/r \pst@number\pst@dima def
\tx@NCCircle \psarc@v end}}
\def\tx@NCBox{NCBox }
\def\ncbox{\pst@object{ncbox}}
\def\ncbox@i{\check@arrow{\ncbox@ii}}
\def\ncbox@ii#1#2{%
\def\pst@linetype{2}%
\nc@object{Closed}{#1}{#2}{.5}{%
tx@Dict begin \psline@iii pop end
\psk@boxheight \psk@boxdepth
\tx@NCBox}}
\def\pcbox{\pst@object{pcbox}}
\def\pcbox@i{\pc@object\ncbox@ii}
\def\tx@NCArcBox{NCArcBox }
\def\psset@boxheight#1{\pst@getlength{#1}\psk@boxheight}
\def\psset@boxdepth#1{\pst@getlength{#1}\psk@boxdepth}
\def\psset@boxsize#1{%
\psset@boxheight{#1}%
\let\psk@boxdepth\psk@boxheight}
\psset@boxsize{.4cm}
\def\ncarcbox{\pst@object{ncarcbox}}
\def\ncarcbox@i{\check@arrow{\ncarcbox@ii}}
\def\ncarcbox@ii#1#2{%
\def\pst@linetype{1}%
\nc@object{Closed}{#1}{#2}{.5}{%
\psk@arcangleA \psk@boxheight \psk@boxdepth \pst@number\pslinearc
\tx@NCArcBox}}
\def\pcarcbox{\pst@object{pcarcbox}}
\def\pcarcbox@i{\pc@object\ncarcbox@ii}
\def\tx@Tfan{Tfan }
\begingroup
\catcode`\:=13
\gdef\pst@activerot{\def:{\string:}}
\endgroup
\def\psset@nrot#1{%
\begingroup
\pst@activerot
\pst@expandafter{\@ifnextchar:{\psset@@nrot}{\psset@@rot}}{#1}\@nil
\global\let\pst@tempg\psk@rot
\endgroup
\let\psk@nrot\pst@tempg}
\def\psset@@nrot:#1\@nil{%
\psset@@rot#1\@nil
\edef\psk@rot{NAngle \ifx\psk@rot\@empty\else\psk@rot add \fi}}
\psset@nrot{0}
\def\tx@LPutCoor{LPutCoor }
\def\tx@LPut{LPut }
\def\psset@npos#1{%
\def\pst@tempa{#1}%
\ifx\pst@tempa\@empty
\def\psk@npos{\npos@default}%
\else
\pst@checknum{#1}\psk@npos
\fi}
\psset@npos{}
\def\ncput{\pst@object{ncput}}
\def\ncput@i{\pst@killglue\pst@makebox{\ncput@ii}}
\def\ncput@ii{%
\begingroup
\use@par
\if@star\pst@starbox\fi
\pst@makesmall\pst@hbox
\pst@rotate\psk@nrot\pst@hbox
\ncput@iii
\endgroup
\pst@shortput}
\def\ncput@iii{%
\leavevmode
\hbox{%
\pst@Verb{%
\pst@nodedict
/t \psk@npos def
\tx@LPut
end
\tx@PutBegin}%
\box\pst@hbox
\pst@Verb{\tx@PutEnd}}}
\def\naput{\pst@object{naput}}
\def\naput@i{\pst@killglue\pst@makebox{\naput@ii{NAngle 90 add}}}
\def\naput@ii#1{%
\begingroup
\use@par
\if@star\pst@starbox\fi
\def\psk@refangle{#1 }%
\let\psk@rot\psk@nrot
\uput@vii
{exch pop add a \tx@PtoC h1 add exch w1 add exch }%
{tx@Dict /NCLW known { NCLW add } if }%
\ncput@iii
\endgroup
\pst@shortput}
\def\nbput{\pst@object{nbput}}
\def\nbput@i{\pst@killglue\pst@makebox{\naput@ii{NAngle 90 sub}}}
\def\psset@tpos#1{%
\pst@checknum{#1}\psk@tpos
\ifdim\psk@tpos \p@<\z@
\def\psk@tpos{.5}%
\@pstrickserr{Bad `tpos' value: `#1'. Must be 0<tpos<1}\@epha
\else
\ifdim\psk@tpos \p@>\p@
\def\psk@tpos{.5}%
\@pstrickserr{Bad `tpos' value: `#1'. Must be 0<tpos<1}\@epha
\fi
\fi}
\psset@tpos{.5}
\def\tvput{\pst@object{tvput}}
\def\tvput@i{\pst@makebox{\psput@tput{H}{1}}}
\def\tlput{\pst@object{tlput}}
\def\tlput@i{\pst@makebox{\psput@tput{H}{true}}}
\def\trput{\pst@object{trput}}
\def\trput@i{\pst@makebox{\psput@tput{H}{false}}}
\def\thput{\pst@object{thput}}
\def\thput@i{\pst@makebox{\psput@tput{V}{1}}}
\def\taput{\pst@object{taput}}
\def\taput@i{\pst@makebox{\psput@tput{V}{true}}}
\def\tbput{\pst@object{tbput}}
\def\tbput@i{\pst@makebox{\psput@tput{V}{false}}}
\def\tx@HPutAdjust{HPutAdjust }
\def\tx@VPutAdjust{VPutAdjust }
\def\psput@tput#1#2{%
\begingroup
\use@par
\pst@tputmakesmall
\leavevmode
\hbox{%
\pst@Verb{%
\pst@nodedict
/t \psk@tpos \pst@tposflip def
tx@NodeDict /HPutPos known
{ #1PutPos }
{ CP /Y ED /X ED /NAngle 0 def /NCLW 0 def }
ifelse
/Sin NAngle sin def
/Cos NAngle cos def
/s \pst@number\pslabelsep NCLW add def
/l \pst@number\pst@dima def
/r \pst@number\pst@dimb def
/h \pst@number\pst@dimc def
/d \pst@number\pst@dimd def
\ifnum1=0#2 \else
/flag #2 def
\csname tx@#1PutAdjust\endcsname
\fi
\tx@LPutCoor
end
\tx@PutBegin}%
\box\pst@hbox
\pst@Verb{\tx@PutEnd}}%
\endgroup
\pst@shortput}
\def\pst@tposflip{}
\def\pst@tputmakesmall{%
\pst@dima=\wd\pst@hbox
\divide\pst@dima 2
\pst@dimg=\psk@href\pst@dimg
\pst@dimb\pst@dima
\advance\pst@dima\pst@dimg % leftsize
\advance\pst@dimb-\pst@dimg % rightsize
\pst@dimd=\psk@vref\relax
\pst@dimc=\ht\pst@hbox
\advance\pst@dimc-\pst@dimd % height
\advance\pst@dimd\dp\pst@hbox % depth
\setbox\pst@hbox=\hbox to\z@{%
\kern-\pst@dima\vbox to\z@{\vss\box\pst@hbox\vskip-\pst@dimd}\hss}}
\def\MakeShortNab#1#2{%
  \def\pst@shortput@nab{%
    \def\pst@tempg{\next}%
    \ifx#1\next
      \let\pst@tempg\naput
    \else
      \ifx#2\next
        \let\pst@tempg\nbput
      \else
        \ifx\@sptoken\next
          \let\pst@tempg\pst@shortput
        \fi
      \fi
    \fi
    \pst@tempg}}
\MakeShortNab{^}{_}
\def\MakeShortTablr#1#2#3#4{%
  \def\pst@shortput@tablr{%
    \def\pst@tempg{\next}%
    \ifx#1\next
      \let\pst@tempg\taput
    \else
      \ifx#2\next
        \let\pst@tempg\tbput
      \else
        \ifx#3\next
          \let\pst@tempg\tlput
        \else
          \ifx#4\next
            \let\pst@tempg\trput
          \else
            \ifx\@sptoken\next
              \let\pst@tempg\pst@shortput
            \fi
          \fi
        \fi
      \fi
    \fi
    \pst@tempg}}
\MakeShortTablr{^}{_}{<}{>}
\def\MakeShortTab#1#2{%
  \def\pst@shortput@tab{%
    \def\pst@tempg{\next}%
    \ifx#1\next
      \def\pst@tempg{%
        \@nameuse{%
          t\ifodd\psk@treemode\ifpstreeflip b\else a\fi
          \else\ifpstreeflip r\else l\fi\fi put}}%
    \else
      \ifx#2\next
        \def\pst@tempg{%
          \@nameuse{%
            t\ifodd\psk@treemode\ifpstreeflip a\else b\fi
            \else\ifpstreeflip l\else r\fi\fi put}}%
      \else
        \ifx\@sptoken\next
          \let\pst@tempg\pst@shortput
        \fi
      \fi
    \fi
    \pst@tempg}}
\MakeShortTab{^}{_}
\def\psset@shortput#1{%
\def\pst@tempg{#1}%
\ifx\pst@tempg\@none
\let\pst@shortput\ignorespaces
\else
\@ifundefined{pst@shortput@#1}%
{\@pstrickserr{Bad short put: `#1'}\@ehpa}%
{\edef\pst@shortput{\noexpand\afterassignment\expandafter\noexpand
\csname pst@shortput@#1\endcsname\noexpand\let\noexpand\next}}%
\fi}
\psset@shortput{none}
\def\lput{\def\pst@par{}\pst@ifstar{\@ifnextchar[{\lput@i}{\lput@ii}}}
\def\lput@i[#1]{\addto@par{ref=#1}\lput@ii}
\def\lput@ii{\@ifnextchar({\lput@iv}{\lput@iii}}
\def\lput@iii#1{\addto@par{nrot=#1}\@ifnextchar({\lput@iv}{\ncput@i}}
\def\lput@iv(#1){\addto@par{npos=#1}\ncput@i}
\def\mput{\def\pst@par{}\pst@ifstar{\@ifnextchar[{\mput@i}{\ncput@i}}}
\def\mput@i[#1]{\addto@par{ref=#1}\ncput@i}
\def\Lput{\def\pst@par{}\pst@ifstar{\@ifnextchar[{\Lput@ii}{\Lput@i}}}
\def\Lput@i#1{\addto@par{labelsep=#1}\Lput@ii}
\def\Lput@ii[#1]{\addto@par{ref={#1}}\@ifnextchar({\Lput@iv}{\Lput@iii}}
\def\Lput@iii#1{\addto@par{nrot={#1}}\@ifnextchar({\Lput@iv}{\Lput@v}}
\def\Lput@iv(#1){\addto@par{npos=#1}\Lput@v}
\def\Lput@v{\pst@killglue\pst@makebox{\Lput@vi}}
\def\Lput@vi{%
\begingroup
\use@par
\if@star\pst@starbox\fi
\Rput@vi
\pst@makesmall\pst@hbox
\pst@rotate\psk@nrot\pst@hbox
\ncput@iii
\endgroup
\pst@shortput}
\def\Mput{\def\pst@par{}\pst@ifstar{\@ifnextchar[{\Mput@ii}{\Mput@i}}}
\def\Mput@i#1{\addto@par{labelsep=#1}\Mput@ii}
\def\Mput@ii[#1]{\addto@par{ref={#1}}\Lput@v}
\def\aput@#1{\def\pst@par{}\pst@ifstar{\@ifnextchar[{\aput@i#1}{\aput@ii#1}}}
\def\aput@i#1[#2]{\addto@par{labelsep=#2}\aput@ii#1}
\def\aput@ii#1{\@ifnextchar({\aput@iv#1}{\aput@iii#1}}
\def\aput@iii#1#2{\addto@par{nrot=#2}\@ifnextchar({\aput@iv#1}{#1}}
\def\aput@iv#1(#2){\addto@par{npos=#2}#1}
\def\aput{\aput@\naput@i}
\def\bput{\aput@\nbput@i}
\def\Aput{\def\pst@par{}\pst@ifstar{\@ifnextchar[{\Aput@i}{\naput@i}}}
\def\Aput@i[#1]{\addto@par{labelsep=#1}\naput@i}
\def\Bput{\def\pst@par{}\pst@ifstar{\@ifnextchar[{\Bput@i}{\nbput@i}}}
\def\Bput@i[#1]{\addto@par{labelsep=#1}\nbput@i}
\def\node@coor#1;#2\@nil{%
\pst@getnode{#1}\pst@tempg
\edef\pst@coor{%
\pst@nodedict
tx@NodeDict \pst@tempg known
{ \pst@tempg load \tx@GetCenter }
{ 0 0 }
ifelse
end }}
\def\Node@coor[#1]#2;#3\@nil{%
\begingroup
\psset{#1}%
\@ifnextchar\bgroup{\Node@@@coor}{\Node@@coor}#2\@nil
\endgroup
\let\pst@coor\pst@tempg}
\def\Node@@coor#1\@nil{%
\pst@getnode{#1}\pst@tempg
\xdef\pst@tempg{%
\pst@nodedict
tx@NodeDict \pst@tempg known
{ \psk@nodesepA \psk@angleA
\pst@tempg load \psk@nodeseptypeA \tx@GetEdge
\psk@offsetA \psk@angleA \tx@AddOffset
\pst@tempg load \tx@GetCenter
3 -1 roll add 3 1 roll add exch }
{ CP }
ifelse
end }}%
\def\Node@@@coor#1{%
\pst@@getcoor{#1}%
\def\psk@angleA{%
\pst@tempg load \tx@GetCenter \pst@coor
3 -1 roll sub 3 1 roll sub neg \tx@Atan}%
\Node@@coor}
\def\nput{\pst@object{nput}}
\def\nput@i#1#2{\pst@killglue\pst@makebox{\nput@ii{#1}{#2}}}
\def\nput@ii#1#2{%
\begingroup
\use@par
\psset@refangle{#1}%
\let\psk@angleA\psk@refangle
\edef\psk@nodesepA{\pst@number\pslabelsep}%
\def\psk@nodeseptypeA{0 }%
\pslabelsep\z@
\uput@vi
\Node@@coor#2\@nil
\let\pst@coor\pst@tempg
\leavevmode
\psput@special\pst@hbox
\endgroup
\ignorespaces}
\newcount\psrow
\newcount\pscol
\newcount\psmatrixcnt
\newskip\psrowsep
\newskip\pscolsep
\def\psset@colsep#1{\pssetlength\pscolsep{#1}}
\def\psset@rowsep#1{\pssetlength\psrowsep{#1}}
\psset@colsep{1.5cm}
\psset@rowsep{1.5cm}
\newif\ifpsmatrix
\let\mscount\@multicnt
\def\psmatrix{%
\begingroup
{\ifnum0=`}\fi % Don't want to expand any &.
\@ifnextchar[{\psmatrix@i}{\ifnum0=`{\fi}{}\psmatrix@ii}}
\def\psmatrix@i[#1]{%
\ifnum0=`{\fi}{}%
\psset{#1}%
\psmatrix@ii}
\def\psmatrix@ii{%
\KillGlue
\edef\psm@beginmath{%
\ifmmode$\m@th\ifinner\textstyle\else\displaystyle\fi\fi}%
\edef\psm@endmath{\ifmmode$\fi}%
\let\\\psm@cr
\advance\psmatrixcnt 1
\def\psm@thenode{M-\the\psmatrixcnt-\the\psrow-\the\pscol}%
\tabskip\z@
\psrow1
\pscol\z@
\psset@shortput{tablr}%
\leavevmode
\vbox\bgroup\halign\bgroup&%
\begingroup
\global\advance\pscol 1
\csname psrowhook\romannumeral\psrow\endcsname
\csname pscolhook\romannumeral\pscol\endcsname
\psm@beginnode##\psm@endnode\endgroup
\cr}
\def\endpsmatrix{%
\crcr\egroup\unskip\egroup
\endgroup}
\def\psm@cr{{\ifnum0=`}\fi\@ifnextchar[{\psm@@cr}{\psm@@@cr{}}}
\def\psm@@cr[#1]{\psm@@@cr{\vskip#1\relax}}
\def\psm@@@cr#1{%
\ifnum0=`{\fi}{}\cr
\noalign{%
\global\advance\psrow 1
\global\pscol\z@
\vskip\psrowsep
#1}}
\def\psm@beginnode{%
\@ifnextchar\psm@endnode
{\let\psm@endnode@i\relax\setbox\pst@hbox=\hbox{}}%
{\pst@object{psm@beginnode}}}
\def\psm@beginnode@i{%
\setbox\pst@hbox=\hbox\bgroup
\psm@beginmath
\begingroup
\ignorespaces}
\def\psm@endnode@i{%
\unskip
\endgroup
\psm@endmath
\egroup
\use@par
\@psttrue}
\def\psm@endnode{%
\@pstfalse
\psm@endnode@i
\ifnum\pscol>1 \hskip\pscolsep \fi
\psk@mnodesize
\hfil
\nodealigntrue
\if@pst
\csname mnode@\psk@mnode\endcsname
\else
\csname mnode@\psk@emnode\endcsname
\fi
\psk@mcol
\psk@@mnodesize}
\def\psspan#1{\mscount#1\relax\loop\ifnum\mscount>\@ne \sp@n\repeat}
\def\psset@name#1{\pst@getnode{#1}\psk@name}
\let\psk@name\relax
\def\psset@mcol#1{%
\ifx r#1\relax
\let\psk@mcol\relax
\else
\ifx l#1\relax
\let\psk@mcol\hfill
\else
\let\psk@mcol\hfil
\fi
\fi}
\psset@mcol{c}
\def\psset@mnodesize#1{%
\pssetlength\pst@dimg{#1}%
\ifdim\pst@dimg<\z@
\let\psk@mnodesize\relax
\let\psk@@mnodesize\relax
\else
\edef\psk@mnodesize{\noexpand\hbox to\number\pst@dimg sp\noexpand\bgroup}%
\let\psk@@mnodesize\egroup
\fi}
\psset@mnodesize{-1pt}
\def\mnode@R{\rnode@iii\Rnode@ii{\psm@thenode}}
\def\mnode@r{\rnode@iii\rnode@iv{\psm@thenode}}
\def\mnode@oval{\ovalnode@ii{\psm@thenode}}
\def\mnode@tri{\trinode@ii{\psm@thenode}}
\def\mnode@dia{\dianode@ii{\psm@thenode}}
\def\mnode@C{{\nodealigntrue\cnode@ii(\z@,\z@){\psk@radius}{\psm@thenode}}}
\def\mnode@f{{\nodealigntrue\fnode@ii(\z@,\z@){\psm@thenode}}}
\def\mnode@circle{\circlenode@ii{\psm@thenode}}
\def\mnode@p{\pnode(\z@,\z@){\psm@thenode}}
\def\mnode@dot{\dotnode@ii(\z@,\z@){\psm@thenode}}
\def\mnode@none{\box\pst@hbox}
\def\psset@mnode#1{%
\@ifundefined{mnode@#1}%
{\@pstrickserr{\string\psmatrix\space node `#1' not defined.}\@ehpa}%
{\edef\psk@mnode{#1}}}
\def\psset@emnode#1{%
\@ifundefined{mnode@#1}%
{\@pstrickserr{\string\psmatrix\space node `#1' not defined.}\@ehpa}%
{\edef\psk@emnode{#1}}}
\psset@mnode{R}
\psset@emnode{none}
\def\nccoil{\pst@object{nccoil}}
\def\nccoil@i{\check@arrow{\nccoil@ii}}
\def\nccoil@ii#1#2{\nc@object{Open}{#1}{#2}{.5}{%
\tx@NCCoor
tx@Dict begin
4 2 roll
\psk@coilwidth \pscoilheight
\psk@coilarmA \psk@coilarmB
\psk@coilaspect \psk@coilinc
\pst@coildict \tx@Coil end
end}}
\def\nczigzag{\pst@object{nczigzag}}
\def\nczigzag@i{\check@arrow{\nczigzag@ii}}
\def\nczigzag@ii#1#2{\nc@object{Open}{#1}{#2}{.5}{%
\tx@NCCoor
tx@Dict begin
4 2 roll
\pscoilheight
\psk@coilwidth
\psk@coilarmA
\psk@coilarmB
\pst@coildict \tx@ZigZag end
\psline@iii
\tx@Line
end}}
\catcode`\@=\TheAtCode\relax
 
\def\ap{'\thinspace}
\def\ds{\displaystyle}
\def\ns{\hspace{-1mm}}

\newcommand{\scez}{\setcounter{equation}{0}}
\newtheorem{lemma}{Lemma}[section]
\newtheorem{theorem}{Theorem}[section]

\newtheorem{proposition}{Proposition}[section]

\newtheorem{problem}{Problem}[section]

\def\be{\begin{equation}}
\def\ee{\end{equation}}
\def\bea{\begin{eqnarray}}
\def\eea{\end{eqnarray}}
\def\beann{\begin{eqnarray*}}
\def\eeann{\end{eqnarray*}}
\def\bsea{\begin{subeqnarray}}
\def\esea{\end{subeqnarray}}
\def\bmat{\left[ \begin{array}}
\def\emat{\end{array} \right]}
\def\bsmat{\left[ \begin{smallmatrix}}
\def\esmat{\end{smallmatrix} \right]}
\def\second{{\prime \prime}}
\def\ap{'\thinspace}
\def\ds{\displaystyle}
\def\nns{\hspace{-.5mm}}%mio
\def\proof{\noindent{\bf{\em Proof:}\ \ }}
\def\QED{\mbox{\rule[0pt]{1.5ex}{1.5ex}}}
\def\endproof{\hspace*{\fill}~\QED\par\endtrivlist\unskip}
\def\endex{\hspace*{\fill}~$\square$\par\endtrivlist\unskip}
\newcommand{\real}{{\mathbb{R}}}
\newcommand{\complex}{{\mathbb{C}}}
\newcommand{\integer}{{\mathbb{Z}}}
\newcommand{\tp}{{^\tra}}
\def\gA{{\cal A}}
\def\gB{{\cal B}}
\def\gC{{\cal C}}
\def\gD{{\cal D}}
\def\gC{{\cal C}}
\def\gF{{\cal F}}
\def\gG{{\cal G}}
\def\gH{{\cal H}}
\def\gI{{\cal I}}
\def\gJ{{\cal J}}
\def\gK{{\cal K}}
\def\gL{{\cal L}}
\def\gM{{\cal M}}
\def\gN{{\cal N}}
\def\gO{{\cal O}}
\def\gP{{\cal P}}
\def\gQ{{\cal Q}}
\def\gR{{\cal R}}
\def\gS{{\cal S}}
\def\gT{{\cal T}}
\def\gU{{\cal U}}
\def\gV{{\cal V}}
\def\gW{{\cal W}}
\def\gX{{\cal X}}
\def\gY{{\cal Y}}
\def\gZ{{\cal Z}}
\def\Ricc{{\bf{{R}}}}
\def\t{\textrm}
\def\second{{\prime \prime}}
\def\tilda{{\!\!\!\!\phantom{P}^\thicksim}}
\def\mtilda{{\!\!\!\!\phantom{P}^{-\thicksim}}}
\def\third{\prime \prime \prime}
\newcommand{\ima}{\operatorname{im}}
\newcommand{\rank}{\operatorname{rank}}
\newcommand{\normrank}{\operatorname{normrank}}
\newcommand{\diag}{\operatorname{diag}}
\newcommand{\defi}{\stackrel{\text{\tiny def}}{=}}
\def\emat{\end{array} \right]}
\def\ssp{\phantom{{\underline g}}}
\def\shp{\phantom{\widehat{l}}}
\def\spp{\phantom{\Big|}}
\newgray{verylightgray}{.80}

\def\tra{{\scalebox{.6}{\thinspace\mbox{T}}}}
\def\ssquare{\tiny{_{\square}}}
\definecolor{Royalblue}{cmyk}{1,0.30,0.2,0.2}
\newcommand{\agu}{\color{Royalblue}}

\begin{document}
\begin{titlepage}
\title{\vspace{15mm}
A Unified Analytical Design Method \\ of Standard Controllers using Inversion Formulae \thanks{Partially supported by the Italian Ministry for Education and Research (MIUR) under PRIN grant n. 20085FFJ2Z).}\vspace{10mm}}
\author{{\large Lorenzo Ntogramatzidis$^\star$, Roberto Zanasi$^\ddagger$ and Stefania Cuoghi$^\ddagger$}\\
       {\small $^\ddagger$DII-Information Engineering Department}\\
        {\small    University of Modena and
Reggio Emilia, Modena, Italy.}\\
       {\small     {\tt roberto.zanasi@unimore.it, stefania.cuoghi@unimore.i}} \\
       {\small     $^\star$Department of Mathematics and Statistics}\\
       {\small     Curtin University, Perth WA, Australia.}\\
       {\small    {\tt L.Ntogramatzidis@curtin.edu.au}}
 }%
\maketitle
\begin{center}
\begin{minipage}{14.2cm}
\begin{center}
\bf Abstract
\end{center}
The aim of this paper is to present a comprehensive range of design
techniques for the synthesis of the standard compensators (Lead and
Lag networks as well as PID controllers) that in the last twenty
years have proved to be of great educational value in a vast number
of undergraduate and postgraduate courses in Control throughout
Italy, but that to-date remain mostly confined within this country.
These techniques hinge upon a set of simple closed-form formulae for
the computation of the parameters of the controller as functions of
the typical specifications introduced in Control courses, i.e., the steady-state
performance, the stability margins and the crossover frequencies.
\end{minipage}
\end{center}
\begin{center}
\begin{minipage}{14.2cm}
\vspace{2mm}
\end{minipage}
\end{center}
\thispagestyle{empty}
\end{titlepage}
\section{Introduction}
The standard compensators presented in every course
and textbook on control systems design belong to two important
families: {\em Lead/Lag networks} and {\em PID controllers}. The
structures of the controllers of these two big sets of compensators
are particularly simple, and this partly justifies the standard
practice of using these compensators as prototype examples to
illustrate the various control synthesis methods in Control courses. However, the importance of these compensator
structures resides also in their relevance in applications. It is
often argued that the 90\% of the compensators used in industry is
made up of PID controllers alone \cite{Astrom-H-01,Dwyer}.

%The study of design and tuning techniques for standard regulators is a cornerstone of both undergraduate and postgraduate courses in Control everywherein the World.

The common trend in both traditional and modern approaches to
Control education is to formulate the feedback control problem as
one in which the design specifications are first expressed using time domain parameters of the response (speed of the response,
overshoot, undershoot, steady-state accuracy, {\em etc}). These requirements are
then transformed into frequency domain specifications (DC gain,
bandwidth, resonant peak, phase and gain margins, crossover
frequencies, {\em etc}). Alternatively -- but less realistically
from a practical perspective -- the design specifications can be
expressed from the very beginning in the frequency domain.

In both situations, the design is effectively carried out using
frequency domain considerations on Bode, Nyquist or -- nowadays less
frequently -- Nichols plots, which constitute different types of
graphical representations of the frequency responses involved in the
control problem. The tuning techniques introduced in the vast
majority of control courses are mainly based on trial-and-error
considerations on these diagrams. For example, when a Lead network
is employed in a feedback control system with the objective of
increasing the phase margin of the loop gain, one can use the
well-known analytical formula that provides the frequency at which
the maximum phase lead is delivered by the network, and impose that
frequency to be equal to the gain crossover frequency of the
uncompensated plant. However, this procedure is not exact, as it
does not take into account the fact that any Lead network with unity
DC gain amplifies the magnitude of the frequency response of the
plant at all (finite) frequencies, and hence the gain crossover
frequency of the loop gain will necessarily be greater than the one
in which the maximum lead is attained. Therefore, the specification
on the phase margin is not exactly met. This suggests that the
problem of placing this frequency can be solved iteratively using
rules of thumb. This is the approach usually taken in the majority
of Control courses and textbooks.
 This renders the synthesis
procedure rather clumsy, and less suitable to be employed for educational purposes.
This is particularly true within the context of written exercises,
not in the least because all the aforementioned plots can only be
drawn with pen and paper only in a very approximate fashion,
especially nowadays when less and less emphasis is given to the
rules for drawing these diagrams as a result of the increasing role
that MATLAB$^{\textrm{\tiny{\textregistered}}}$ has to the same
purpose.
 It is very difficult to construct a written exercise, or test, of exam, in
which the control design problem consists in the exploitation of
graphical techniques to compute the parameters of the desired
compensator. Another
consequence of the clumsiness associated with the classic
trial-and-error design method is the fact that this procedure is
difficult to automate into an algorithm
that can be used for educational purposes (and for the same reason it is also unsuited to be used as a self-tuning
strategy).

%An equally important drawback of these techniques is the fact that
%they appear to be highly unsuitable for didactical purposes.  The situation is different when a student has access to
%an automatic computing environment such as
%MATLAB$^{\textrm{\tiny{\textregistered}}}$.
%In fact, nowadays less and less emphasis is given in Control courses to the teaching of all the rules that

In this paper, we present an alternative methodology that can be
successfully employed both in an educational and in a practical
context to carry out the design of a standard compensator given
standard control system specifications such as steady-state
performance, gain and phase crossover frequencies, phase and gain
margins. This method is based on a set of very simple closed-form
formulae, known as {\em Inversion Formulae}, which deliver the
parameters of the compensator as an {\em explicit} function of the
specifications. These formulae first appeared for generic
first-order compensators in \cite{Phillips-85}, and their geometric
interpretation in the context of control feedback design was
explained in \cite{Zanasi-M-76_1}. Surprisingly, the
 pioneering paper \cite{Phillips-85}, which gives an extremely
powerful tool for the design of standard compensators, has never
been cited in the literature, and this in part explains why this
method is still relatively unknown to the wider scientific
community. As such, its potential as a precise and meaningful tool
in Control education is still to be fully examined.
%its potential to lead to a much more precise and meaningful educational tool.
A significant exception which is worth mentioning is the Italian
Control literature. In the Technical Report \cite{Policastro-Z-82}
the procedure presented in \cite{Phillips-85} had already been
outlined for Lead and Lag networks. This technique also appeared in
the Italian control textbook \cite{Marro-04}. This undergraduate
textbook has been by far the most utilised one in University courses
and technical secondary institution ({\em Istituti Tecnici}) courses
throughout Italy over the past twenty years. Due to its popularity,
the same technique has later appeared in other University textbooks
in Italy, see e.g. \cite{Marro-98} and \cite{Ferrante-LV-00}. In the
latter, a hint on how to adapt this technique to PID controllers was
also presented. However, the success of this technique for
educational purposes has so far remained confined within the Italian
Control literature.

The aim of this paper is to present this technique in the most
comprehensive way possible, to make its potential in control systems
design education clear to the wider scientific community.
% We aim to derive a unified framework for the employment of these formulae

The educational value of the method outlined in this paper is motivated by the following facts.
\begin{enumerate}
\item
The entire synthesis procedure can be carried out by pen, paper and
a scientific calculator; it is therefore very suitable to be employed in
all forms of written questions and exercises;
\item
The synthesis procedure forces the students to follows the classical
order of taking into account the steady state specifications first,
and then to design the remaining part of the compensator;
\item
Even though the synthesis methodology described here can be carried
out by pen and paper, this technique has also an important graphical
counterpart. In other words, it is shown that the Inversion Formulae
enable the control system design problem to be solved analytically
with pen and paper, or  graphically on Nyquist, Bode or Nichols
plots (without necessarily using trial-and-error or iterative
procedures), thus retaining important links to other parts of a
programme of a course of Control, \cite{Ntogramatzidis-112};
\item
Unlike the traditional design methodologies, the feasibility of the design procedure can be checked {\em a
priori}. Furthermore, once the Bode gain of the compensator is
computed from the steady-state requirements, very simple
considerations can lead students to the selection of the most
suitable type of compensator to be employed; %Considerations on Bode
%and Nyquist plots can be used to this end; however, only a
%qualitative plot is needed to this purpose, and the precision with
%which the solution is calculated does not depend on the precision
%with which these diagrams are drawn;
\item
The mathematical tools that are needed to explain the method are
basic notions of trigonometry and complex numbers. Hence, the use of
Inversion Formulae reinforces the use of manipulations of complex
numbers which is crucial in control systems education;
\item
The situations in which some of the parameters of the compensator
turn out to be positive can be fruitfully linked to important
considerations on the shape of the Bode plot of the compensator;
\item
The method based on the Inversion Formulae can be implemented as an
extremely simple algorithm, for example using
MATLAB$^{\textrm{\tiny{\textregistered}}}$. An example will be
presented in this paper;
\item
For the most part, there is a tendency of Control courses and
textbooks to neglect the synthesis techniques for richer compensator
structures such as the Lead-Lag network. The Inversion Formulae
enable these compensators to be addressed without a significant
increase in the design complexity. This is an important advantage,
because Lead-Lag networks offer additional flexibility with respect
to standard Lead and Lag networks, that results in the ability to
satisfy further specifications or constraints.
\end{enumerate}

In this paper, we present the design technique based on the
Inversion Formulae by first presenting the feedback control problem
in the way it is usually introduced in undergraduate and
postgraduate Control courses and textbooks. Lead and Lag networks
will be the first compensator structures to be considered. It will
be shown how simple considerations on the plant transfer function
and on the specifications of the problem can guide students to the
choice of the correct network to employ. In the second part of the
paper, PID controllers will be introduced. The design approach is
similar in spirit to the one for Lead-Lag networks, but the way
steady-state specifications are accommodated in these two scenarios
are slightly different. Indeed, the formulae that deliver the
parameters of the PID controller depend on the type of stead-state
specification, as also shown in \cite{Ntogramatzidis-F-11}. The
understanding of such difference is a crucial aspect in the
understanding of the problem of steady-state specifications in a
control feedback design problem.

In this paper we give great emphasis to the numerical examples, because these show what we believe is the most important feature of the Inversion Formulae in Control education, i.e., in the possibility of devising simple and at the same time complete and educationally relevant written exercises that have the potential to guide students through all stages of the compensator design process. To stress the potential offered by these formulae in all kinds of written exercises we illustrate the solutions of the numerical problems proposed here in a closed-form. However, we also show that this analytical method also has a fundamental graphical counterpart that adds a further dimension to the learning experience of the synthesis of standard compensators as highlighted in \cite{Ntogramatzidis-112}.

\section{Formulation of the control problem}
\begin{figure}[htb]
\centering
\psfrag{R}[]{\small $R(s)$}
\psfrag{E}[]{\small $\,$}
\psfrag{G}[]{\small $Y(s)$}
\psfrag{E}[]{\small $E(s)$}
\psfrag{I}[]{\tiny $+$}
\psfrag{J}[]{\tiny $-$}
\psfrag{U}[]{\small $U(s)$}
\psfrag{H}[]{\small $\ssp \,\, H(s) \ssp$}
\psfrag{B}[]{\small $\ssp \,\, C(s) \ssp$}
\psfrag{A}[]{\small $\ssp \,\, G(s)  \ssp$}
\includegraphics[width=2.5in]{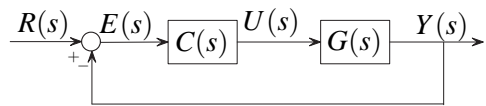}
\caption{Classic feedback control architecture.}
\label{fig1}
\end{figure}

 Consider the classic feedback control architecture in Figure \ref{fig1}, where $G(s)$ is the transfer function of the plant, which is assumed to be stable.
%%%%%%%%%%%%%%%%%%%%%%%%%%%%%%%%%%%%
%%%%%%%   FIGURE   %%%%%%%%%%%%%%%%%
%%%%%%%%%%%%%%%%%%%%%%%%%%%%%%%%%%%%
%
In Figure \ref{fig1}, the symbols $R(s)$, $U(s)$ and $Y(s)$
respectively represent the Laplace transforms of the reference
signal $r(t)$, of the control input $u(t)$ and of the controlled
output $y(t)$. Let $E(s)$ represent the Laplace transform of the
tracking error $e(t)\stackrel{\text{\tiny def}}{=}r(t)-y(t)$. The
first and most basic control problem that we consider is the one
that aims at satisfying standard specifications on the steady-state
performance, on the phase margin and on the gain crossover
frequency. To express these specifications mathematically, we define the
loop gain transfer function as the product $L(s)
\stackrel{\text{\tiny def}}{=} {C}(s)\,{G}(s)$.
When $L(s)$ is strictly proper and the
polar plot of $L(j \omega)$ for $\omega \ge 0$ has a single intersection with the unit
circle and the negative real semiaxis (except for the
trivial intersection at the origin as $\omega \to \infty$), the gain and phase margins are well defined, and ensure that
the polar plot of $L(j \omega)$ does not encircle the critical point
$-1$ in view of the simplified version of the Nyquist criterion,
\cite{Franklin-PE-06}. We denote by $\omega_g$ the gain crossover
frequency, i.e., the frequency at which the polar plot of $L(j
\omega)$ intersects the unit circle. Hence, $\omega_g$ is such that
$|L(j \omega_g)| = 1$, and the phase margin is defined as the angle
$\textrm{PM} \stackrel{\text{\tiny def}}{=} \textrm{arg}\,L(j
\omega_g)+\pi$. Similarly,  we denote by $\omega_p$ the phase
crossover frequency, i.e., the frequency at which the polar plot of
$L(j \omega)$ intersects the negative real half-axis. As such,
$\omega_p$ is such that $\textrm{arg}\,L(j \omega_p) = -\pi$, and
the gain margin is defined as $\textrm{GM} \stackrel{\text{\tiny
def}}{=} 1/ |L(j \omega_p)|$.\\[-2mm]

\begin{problem}
\label{pro1} Find a controller $C(s)$ such that the steady-state
requirements on the tracking error $e(t)$ are satisfied, and such
that the gain crossover frequency and the phase margin of the loop
gain transfer function $L(s)$ are $\omega_g$ and $\textrm{PM}$,
respectively, i.e., such that \bea |L(j \omega_g)| = 1 \quad
\textrm{and} \quad \textrm{arg}\,L(j \omega_g) = \textrm{PM}-\pi.
\label{L1} \eea
\end{problem}

An alternative control problem can be formulated by specifying the
gain margin and the phase crossover frequency:
\begin{figure}[tbp]
  \centering
  \psfrag{0}[lb][l][0.8]{$0$}
  \psfrag{-1}[br][br][0.9]{$-1$}
  \psfrag{M}[c][c][1]{$\frac{1}{GM}$}
  \psfrag{Im}[r][r][0.9 ]{$Im$}
  \psfrag{Re}[r][r][0.9 ]{$Re$}
  \psfrag{P}[tr][tr][0.9 ]{$PM$}
  \psfrag{wg}[r][r][0.8]{$\omega_g$}
  \psfrag{wp}[r][r][0.8]{$\omega_p$}
  \psfrag{L}[l][l][0.9]{$\hspace{1mm}L(j\omega)$}
  \includegraphics[width=7.3cm]{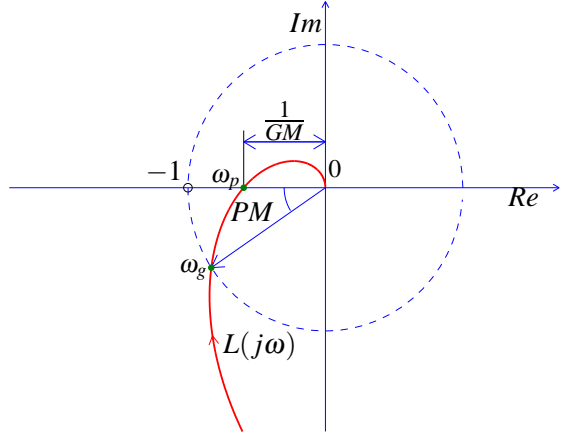}
  \caption{Design specifications: gain margin $\textrm{GM}$, phase margin $\textrm{PM}$, gain crossover frequency $\omega_g$ and phase crossover frequency $\omega_p$.}\label{Design_specifications_Gm_Pm}
\end{figure}
\ \\[-2mm]

\begin{problem}
\label{pro2} Find $C(s)$ such that the steady-state requirements on
the tracking error are satisfied, and such that the phase crossover
frequency and the gain margin of $L(s)$ are $\omega_p$ and
$\textrm{GM}$, respectively, i.e., such that \bea
 |L(j \omega_p)| = \textrm{GM}^{-1} \quad \textrm{and} \quad  \textrm{arg}\,L(j \omega_p) =-\pi. \label{L2}
\eea
\end{problem}

In some cases, the compensators with a richer dynamic structure will
allow an additional degree of freedom to be exploited to the end of
satisfying a further specification. In these cases, the control
problem considered here is the one in which, in addition to the
steady-state specification, the gain crossover frequency, the phase
and the gain margin are imposed. This is the case of Lead-Lag
networks and of PID controllers in which the steady-state
performance requirements do not lead to a constraint on the Bode
gain.\\[-2mm]

\begin{problem}
\label{pro3} Find a controller $C(s)$ that meets the steady-state
requirements, and such that the gain crossover frequency, the phase
margin and the gain margin of the $L(s)$ are $\omega_g$,
$\textrm{PM}$ and $\textrm{GM}$, respectively. In other words,
$C(s)$ must guarantee that a frequency $\omega_p>0$ exists such that
(\ref{L1}) and (\ref{L2}) hold.
\end{problem}

The first step of the design procedure consists in writing the transfer function $C(s)$ of
the compensator as the product of a constant $K$, that is determined
by imposing the steady-state requirements, by the transfer function
$\bar{C}(s)$ with unity DC gain.

\begin{figure}[htb]
\centering
\psfrag{R}[]{\small $R(s)$}
\psfrag{E}[]{\small $\,$}
\psfrag{G}[]{\small $Y(s)$}
\psfrag{I}[]{\tiny $+$}
\psfrag{J}[]{\tiny $-$}
\psfrag{U}[]{\small $U(s)$}
\psfrag{E}[]{\small $E(s)$}
\psfrag{H}[]{\small $\ssp \,\, H(s) \ssp$}
\psfrag{B}[]{\small $\ssp \,\,\bar{C}(s)\ssp$}
\psfrag{A}[]{\small $\ssp \,\, K \cdot G(s)  \ssp$}
\includegraphics[width=2.5in]{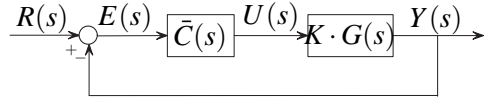}
\caption{Modified feedback control architecture with a unity DC gain network.}
\label{fig2}
\end{figure}

% As is well known from classical control theory, in the unity feedback case, or when the DC gain of $H(s)$ is equal to one, steady-state specifications that Lead to the sharp assignment of the steady-state error to a given non-zero constant are such that the DC gain of the network is fixed. More precisely, if we assign the position error $e_{\infty}^p$ for type-0 plants, the velocity error $e_{\infty}^v$ for type-1 plants, or the acceleration error $e_{\infty}^a$ for type-2 plants, to a given non-zero constant, the DC gain of the network is determined.
Since the term $K$ is known after the steady-state constraints have
been imposed, it can be considered as being part of the plant, see
Figure \ref{fig2}.
%%%%%%%%%%%%%%%%%%%%%%%%%%%%%%%%%%%%
%%%%%%%   FIGURE   %%%%%%%%%%%%%%%%%
%%%%%%%%%%%%%%%%%%%%%%%%%%%%%%%%%%%%
%
Let us define $\bar{G}(s)\stackrel{\text{\tiny def}}{=} K\,G(s)$.
 In order to compute the parameters of the compensator, we write $\bar{G}(j \omega)$ and $\bar{C}(j \omega)$ in polar form as
\beann
\bar{G}(j \omega)= |\bar{G}(j \omega)|\,e^{\,j\,\textrm{arg}\,\bar{G}(j \omega)}, \quad   \bar{C}_{\rm \,}(j \omega)=M(\omega)\,e^{\,j\,\varphi(\omega)}.
\eeann

The loop gain frequency response can be written as $L(j \omega)=
|\bar{G}(j \omega)|\,M(\omega)\,e^{\,j\,\left(\textrm{arg}\,{G}(j
\omega)+\varphi(\omega)\right)}$. Consider Problem \ref{pro1}. From
(\ref{L1}) we find
%If the gain crossover frequency $\omega_g$ and the phase margin $\textrm{PM}$ of the loop gain transfer function $L(s)$ are assigned, by (\ref{L1}-\ref{L2}) it is found that $|L(j \omega_g)| = 1$ and $\textrm{PM}  = \pi+\textrm{arg}\,L(j \omega_g)$ must be satisfied. These two equations can be written as
\begin{description}
\item{\em i-a) \hspace{2mm}} $M_g= {1}/{\phantom{\Big|}|\bar{G}(j \omega_g)|\phantom{\Big|}}$,
\item{\em ii-a) \hspace{2mm}} $\varphi_g=\textrm{PM}-\pi-\textrm{arg}\,\bar{G}(j \omega_g)$,
\end{description}
where $M_g\stackrel{\text{\tiny def}}{=}M(\omega_g)$ and
$\varphi_g\stackrel{\text{\tiny def}}{=}\varphi(\omega_g)$. At this
point, since $M_g$ and $\varphi_g$ are known, by solving the
equation \bea \label{inversion} \bar{C}(j
\omega_g)=M_g\,e^{j\,\varphi_g} \eea via the so-called {\em
Inversion Formulae} we find all the remaining parameters of the
compensator. In the case of Problem \ref{pro2}, the imposition of
(\ref{L2}) leads to
\begin{description}
\item{\em  i-b) \hspace{2mm}} $M_p=  {1}/({\phantom{\Big|}\textrm{GM}\,|\bar{G}(j \omega_p)|\phantom{\Big|}})$,
\item{\em  ii-b) \hspace{2mm}} $\varphi_p=-\pi-\textrm{arg}\,\bar{G}(j \omega_p)$,
\end{description}
and the equation to be solved has the same structure of
(\ref{inversion}), with $M_p$, $\varphi_p$ and $\omega_p$ instead of
$M_g$, $\varphi_g$ and $\omega_g$, respectively.

\subsection{Standard compensators} \label{Standard_compensators}

The families of compensators that are considered in this paper are the
phase-correction networks (Lead, Lag and Lead-Lag) and the PID controllers. These are the two
most studied and utilised types of compensators, and are those that are introduced
with no exceptions in all undergraduate and postgraduate textbooks
of control feedback design:

\subsubsection{Phase-Correction Networks}
\begin{itemize}
\item Lead network: $C_{\rm Lead}(s)=K\,\dfrac{1+\tau\,s}{1+\alpha\,\tau\,s}$ \\[0mm]
\item Lag network: $C_{\rm Lag}(s)=K\,\dfrac{1+\alpha\,\tau\,s}{1+\tau\,s}$\\[0mm]
\item Lead-Lag network: $C_{\rm LL}(s)= K   \,\dfrac{(1+\tau_1\,s)(1+\tau_2\,s)}{(1+\alpha\,\tau_1\,s)(1+\frac{\tau_2}{\alpha}\,s)}$
\end{itemize}
where $\alpha \in (0,1)$ and $\tau,\tau_1,\tau_2>0$. The transfer
function of the Lead-Lag network used in this paper generalises the
one given above, because it includes the case of complex conjugate
poles and zeros:
\begin{itemize}
\item Lead-Lag network with complex poles and zeros: \\ $C_{\rm LL}^\prime(s)= K\,\dfrac{s^2+2\,\zeta_1\,\omega_n\,s+\omega_n^2}{s^2+2\,\zeta_2\,\omega_n\,s+\omega_n^2}$,
with $\zeta_1, \,\zeta_2>0$ and $\omega_n>0$. The Lead-Lag network
$C_{\rm LL}(s)$ can always be written as $C_{\rm LL}^\prime(s)$ by
setting \bea \zeta_1 =
\dfrac{\tau_1+\tau_2}{2\,\sqrt{\tau_1\,\tau_2}}, \hspace{2mm}
\zeta_2 =
\dfrac{\alpha\,\tau_1+\frac{\tau_2}{\alpha}}{2\,\sqrt{\tau_1\,\tau_2}},
\hspace{2mm} \omega_n = \dfrac{1}{\sqrt{\tau_1\,\tau_2}}.
\label{legame} \eea Conversely, $C_{\rm LL}^\prime(s)$ can be
written as in $C_{\rm LL}(s)$ if and only if $\zeta_1>1$ and
$\zeta_2>1$. In this case, by defining $\hat{\zeta}_1^{\pm} :=
\zeta_1 \pm \sqrt{\zeta_1^2-1}$ and $\hat{\zeta}_2^{\pm} := \zeta_2
\pm \sqrt{\zeta_2^2-1}$, $C_{\rm LL}^\prime(s)$ can be written as in
$C_{\rm LL}(s)$ with: \beann
&& \hspace{-0.4cm} \alpha=\hat{\zeta}_2^{-}/\hat{\zeta}_1^{-}, \hspace{2mm} \tau_1=\hat{\zeta}_1^{-}/\omega_n, \hspace{2mm} \tau_2=\hat{\zeta}_1^{+}/\omega_n \hspace{2mm} \textrm{if} \hspace{2mm} \zeta_1<\zeta_2; \\
&& \hspace{-0.4cm}  \alpha=\hat{\zeta}_2^{\pm}/\hat{\zeta}_1^{+},
\hspace{2mm} \tau_1=\hat{\zeta}_1^{+}/\omega_n, \hspace{2mm}
\tau_2=\hat{\zeta}_1^{-}/\omega_n  \hspace{2mm} \textrm{if}
\hspace{2mm}\zeta_1>\zeta_2. \eeann
%\end{itemize}
\end{itemize}
\subsubsection{PID controllers}
\begin{itemize}
\item PID controller: $C_{\rm PID}(s)=K_p\,\left( 1+\dfrac{1}{T_i\,s}+T_d\,s \right)$
\item PI controller: $C_{\rm PI}(s)=K_p\,\left(1+\dfrac{1}{T_i\,s} \right)$
\item PD controller: $C_{\rm PD}(s)=K_p\,\left( 1+T_d\,s \right)$
\end{itemize}
with $K_p,T_i,T_d> 0$. In addition to these controllers, sometimes
the proper versions of the PID and PD controllers are also
introduced. The second one is basically equivalent to a Lead
network. These more complex structures will not be considered here.
For details on these structures, see \cite{Ntogramatzidis-F-11}.

\section{Lead, Lag and Lead-Lag networks}
We begin by considering compensators which belong to the family of
Lead and Lag networks. The first step consists in the computation of
the DC gain $K$ of the phase-correction network, using the
steady-state specifications. For phase-correction networks we can
isolate the static gain by writing $C_{\rm Lead}(s)=K\,\bar{C}_{\rm
Lead}(s)$, $C_{\rm Lag}(s)=K\,\bar{C}_{\rm Lag}(s)$ and $C_{\rm
LL}(s)=K\,\bar{C}_{\rm Lead-Lag}(s)$, where $\bar{C}_{\rm Lead}(s)$,
$\bar{C}_{\rm Lag}(s)$ and $\bar{C}_{\rm LL}(s)$ have unity DC gain.
As is well known from classical control theory, in the unity
feedback case, steady-state specifications that lead to the sharp
assignment of the steady-state error to a given non-zero constant
are such that the DC gain of the network is fixed. More precisely,
if we assign the position error $e_p$ for type-0 plants, the
velocity error $e_v$ for type-1 plants, or the acceleration error
$e_a$ for type-2 plants, to a given non-zero constant, the DC gain
of the network is
 determined.

When our aim is to solve Problems \ref{pro1} using a Lead, Lag or
Lead-Lag network (that from now on will be considered with unity DC
gain), simple standard considerations on the frequency response of
these compensators suggest that the choice of the type of
compensator to be employed can be made according to the following
 table.
 \begin{table}[!hbp]
\begin{center}
\vspace{.2cm}
\begin{tabular}{| c || c | c |}
\hline
$\;$  & \phantom{$\Big|$} $M_g>1$ \phantom{$\Big|$} &  $M_g<1$  \\
\hline \hline
$\varphi_g \in \left(-\frac{\pi}{2},0\right)$ & \hspace{-4mm} \phantom{$\Big|$} Lead-Lag ($\tau_2 > \tau_1$) \phantom{$\Big|$}  \hspace{-4mm}  &   \hspace{-4mm}  \phantom{$\Big|$} Lag or Lead-Lag ($\tau_1 > \tau_2$)  \phantom{$\Big|$}  \hspace{-4mm} \\
\hline
$\varphi_g \in \left(0, \frac{\pi}{2}\right)$   &  \hspace{-4mm}  \phantom{$\Big|$} Lead or Lead-Lag ($\tau_2 > \tau_1$) $\;$\phantom{$\Big|$}  \hspace{-4mm} &  \hspace{-4mm}  Lead-Lag ($\tau_1 > \tau_2$)  \hspace{-4mm}   \\
\hline
\end{tabular}
\end{center}
\caption{Use of phase-correction networks}
\label{table1}
\end{table}

The considerations that emerge from this table result from any of
the graphical and analytical methods to characterise the frequency
response of the compensator. For example, the Bode plot of the
magnitude of a Lead network shows that such compensator amplifies
the magnitude of the plant at all finite frequencies. Therefore, the
gain crossover frequency of the loop gain to be selected in order
for a Lead network to solve the problem must necessarily be greater
than the gain crossover frequency of the uncompensated system. That
is, the magnitude of the uncompensated system at the frequency that
we want to select as the gain crossover frequency of the loop gain
must be smaller than $1$, so that its inverse (which is $M_g$) must
be greater than $1$. Similar considerations on the Bode plot (or on
the Nyquist or Nichols plots) can be used to justify the other
entries in the table above, and can guide students (and engineers)
towards the choice of the correct type of compensator.

Notice that Table \ref{table1} also gives a reason why usually
Lead-Lag networks are defined only in the case in which
$\tau_1>\tau_2$. In fact, when $\tau_2>\tau_1$, if $\varphi_g \in
\left(0, \frac{\pi}{2}\right)$ Problem \ref{pro1} can be solved
using simply a Lead network, while if $\varphi_g \in
\left(-\frac{\pi}{2},0\right)$ a proportional controller (with gain
equal to $M_g$) can even provide a phase margin greater than
$\textrm{PM}$. However, because of the way Problem \ref{pro1} has
been formulated, it is more natural to also consider the case
$\tau_2>\tau_1$. Moreover, even if in some situations a Lead or a
Lag network can be used instead of a Lead-Lag network to satisfy
specifications on the phase margin and gain crossover frequency, a
Lead-Lag network still presents the advantage of allowing a further
parameter (such as the gain margin) to be assigned as well.

 In order to solve Problems \ref{pro1} and \ref{pro2} for all types of phase-correction networks, we use the following simple lemma.

 \begin{lemma}
 \label{lem}
 Let $P,Q \in \real$, $M \in \real_+$ and $\varphi \in (-\frac{\pi}{2},\frac{\pi}{2})$. Consider the following equation
 \bea
 \label{eq}
 \dfrac{1+j\,P}{1+j\,Q}=M\,e^{j\,\varphi}.
 \eea
 Solving (\ref{eq}) with respect to $P$ and $Q$ yields \\[-2mm]
% \begin{center}
%\psframebox[linecolor=black]{\parbox{5.7cm}{
%\psframebox[linecolor=black]{\parbox{5.43cm}{
 \beann
 P = \dfrac{M-\cos \varphi}{\sin \varphi} \quad
 Q = \dfrac{M\,\cos \varphi-1}{M\,\sin \varphi}
 \eeann
% }} }}
%\end{center}
 \end{lemma}
 \ \\
The proof of Lemma \ref{lem} follows straightforwardly  by equating
the real and imaginary parts of (\ref{eq}) once $M\,e^{j\,\varphi}$
is expressed as $M\,(\cos \varphi+j\,\sin \varphi)$, see also
\cite{Phillips-85}. A geometric proof of the same result can be
found in
 \cite{Zanasi-M-76_1}.

Lemma \ref{lem} is the result that allows the parameters of the phase-correction network to be computed in closed form.  \\

 %\begin{description}
 %\item
 {\bf Lead network:} Equation (\ref{inversion}) with $\bar{C}(j \omega_g)=\bar{C}_{\rm Lead}(j \omega_g)$ is solvable in $\alpha \in (0,1)$ and $\tau>0$ if and only if
 \bea
\label{conditionsLead}
0 < \varphi_g < \frac{\pi}{2}  \quad \textrm{and} \quad M_g>\frac{1}{\cos \varphi_g}.
\eea
If (\ref{conditionsLead}) is satisfied, the solution of (\ref{inversion}) with $\bar{C}(j \omega_g)=\bar{C}_{\rm Lead}(j \omega_g)$ is given by
\bea
\label{inversionLead}
\alpha = \frac{M_g \,\cos \varphi_g-1 }{M_g\,( M_g- \cos \varphi_g )} \quad {\rm and} \quad \tau=\frac{M_g- \cos \varphi_g}{\omega_g\,\sin \varphi_g}.
\eea
 Eqs. (\ref{inversionLead}) are called {\em Inversion Formulae for the Lead network}.
This result is a consequence of Lemma \ref{lem}, with
$P=\tau\,\omega_g$ and $Q=\alpha\,\tau\,\omega_g$. Conditions
(\ref{conditionsLead}) ensure that $\tau>0$ and $\alpha \in (0,1)$.
These conditions can be also written as $M_g > 1$ and $0 <
\varphi_g< \arccos ({1}/{M_g})$. It is of significant educational
value to show the link between the solvability of
(\ref{conditionsLead}) and the dynamic characteristics of the Lead
network: the Bode plot of the magnitude of a Lead network shows that
the effect of this network on the polar plot of the plant is to
amplify the magnitude for all non-zero frequencies: this means that
the gain crossover frequency of the loop gain transfer function is
greater than that of the plant alone. Therefore, it is essential that
the gain crossover frequency to be chosen for the loop gain must be
greater than the one of $\bar{G}(s)$: in case it is not, a Lead
network achieving that goal does not exist. This fact is in line
with the fact that $M_g>1$.

In the case of Problem \ref{pro2}, the solution of (\ref{inversion})
with $\bar{C}(j \omega_p)=\bar{C}_{\rm Lead}(j \omega_p)$ is given
by the same equations written above with $M_p$, $\varphi_p$ and
$\omega_p$ instead of $M_g$, $\varphi_g$ and
 $\omega_g$.\\
%

  %\itemù
  {\bf Lag network:} Equation (\ref{inversion}) with $\bar{C}(j \omega_g)=\bar{C}_{\rm Lag}(j \omega_g)$ is solvable in $\alpha \in (0,1)$ and $\tau>0$ if and only if
  \bea
\label{conditionsLag} -\frac{\pi}{2}< \varphi_g < 0  \quad
\textrm{and} \quad M_g< {\cos \varphi_g}. \eea If
(\ref{conditionsLag}) are satisfied, the solution of
(\ref{inversion}) with $\bar{C}(j \omega_g)=\bar{C}_{\rm Lag}(j
\omega_g)$ is given by
 \bea
\label{inversionLag}
\alpha = \frac{M_g \,(\cos \varphi_g-M_g) }{1- M_g\,\cos \varphi_g } \quad {\rm and} \quad \tau=\frac{M_g\,\cos \varphi_g-1}{\omega_g\,M_g\,\sin \varphi_g},
 \eea
which are called \emph{Inversion Formulae for the Lag network}. This
result follows from Lemma \ref{lem}, with $P=\alpha\,\tau\,\omega_g$
and $Q=\tau\,\omega_g$. Conditions (\ref{conditionsLag}) can be also
written as $M_g < 1$ and $-\arccos M_g < \varphi_g < 0$. \\

  %\itemù
{\bf Lead-Lag network:} As already observed, in this case the
compensator has an additional parameter that can be exploited to
satisfy a further requirement other than the gain crossover
frequency and the phase margin. One can, for example, assign the
value of a parameter (a damping ratio or the natural frequency) and
then solve for the other two. However, a more interesting problem to
be solved in this case is Problem \ref{pro3}. Let us first consider
the Lead-Lag network $\bar{C}_{\rm LL}^\prime(s)$ with complex poles
and zeros and with unity DC gain. Its frequency response can be
written for $\omega \neq \omega_n$
  as
  \beann
\bar{C}_{\rm LL}^\prime(j \,\omega)= \dfrac{1\!+\!j
\,P(\omega)}{1\!+\!j \, Q(\omega)}, \hspace{2mm} P(\omega)=
\dfrac{2\,\zeta_1\,\omega\,\omega_n}{\omega_n^2-\omega^2},
\hspace{2mm}
Q(\omega)=\dfrac{2\,\zeta_2\,\omega\,\omega_n}{\omega_n^2-\omega^2}.
  \eeann
As such, when we want to assign the gain crossover frequency and the
phase margin, we need to solve (\ref{eq}) with $P=P_g=P(\omega_g)$,
$Q=Q_g=Q(\omega_g)$, $M=M_g$ and $\varphi=\varphi_g$. Similarly,
when we want to assign the phase crossover frequency and the phase
margin, we need to solve (\ref{eq}) with $P=P(\omega_p)$,
$Q=Q(\omega_p)$, $M=M_p$ and $\varphi=\varphi_p$. As such, in order
to solve Problem \ref{pro3} we must solve
  \bea
  \label{eqs}
  \left. \begin{array}{rclrcl}
  P_g \ns&\ns = \ns&\ns \dfrac{2\,\zeta_1\,\omega_g\,\omega_n}{\omega_n^2-\omega_g^2} \ns&\ns \qquad  Q_g \ns&\ns = \ns&\ns  \dfrac{2\,\zeta_2\,\omega_g\,\omega_n}{\omega_n^2-\omega_g^2}  \\
  P_p \ns\ns&\ns\ns = \ns\ns&\ns\ns \dfrac{2\,\zeta_1\,\omega_p\,\omega_n}{\omega_n^2-\omega_p^2} \ns&\ns \qquad
  Q_p\ns&\ns   = \ns&\ns \dfrac{2\,\zeta_2\,\omega_p\,\omega_n}{\omega_n^2-\omega_p^2},
  \end{array} \right.
  \eea
in which $P_g$ and $Q_g$ are completely assigned by the
specifications, whereas $P_p$ and $Q_p$ are functions of $\omega_p$
which is not assigned.
  From $\zeta_1/\zeta_2=P_g/P_p=Q_g/Q_p$ we obtain
 \bea
\label{const}
\frac{M_g-\cos \varphi_g}{\cos \varphi_g-\dfrac{1}{M_g}}=\frac{M_p-\cos \varphi_p}{\cos \varphi_p-\dfrac{1}{M_p}},
 \eea
which is an equation in the unknown $\omega_p$. If $G(s)$ is a
rational function of $s \in \complex$, it is a simple exercise of
trigonometry to verify that (\ref{const}) is a polynomial equation
in $\omega_p$, and therefore it is easy to derive {\em all} it
solutions in closed form, whenever the degree is lower or equal to
5, or numerically. Using (\ref{eqs}) we find the parameters
 \beann
\zeta_1 \ns&\ns = \ns&\ns \dfrac{\omega_g^2-\omega_p^2}{2\,\Phi_2}\,\,\sqrt{\dfrac{\Phi_2}{\omega_g\,\omega_p\,\Phi_1}}   \\
  \zeta_2 \ns&\ns = \ns&\ns \dfrac{\omega_g^2-\omega_p^2}{2\,\Psi_2}\,\,\sqrt{\dfrac{\Psi_2}{\omega_g\,\omega_p\,\Psi_1}}   \\
\omega_n \ns&\ns = \ns&\ns \sqrt{\omega_g\,\omega_p\, \frac{\Phi_1}{\Phi_2}}=\sqrt{\omega_g\,\omega_p\, \frac{\Psi_1}{\Psi_2}}
 \eeann
where $\Phi_1=\omega_g\,P_p^{-1}-\omega_p\,P_g^{-1}$,
$\Phi_2=\omega_p\,P_p^{-1}-\omega_g\,P_g^{-1}$,
$\Psi_1=\omega_g\,Q_p^{-1}-\omega_p\,Q_g^{-1}$ and
$\Psi_2=\omega_p\,Q_p^{-1}-\omega_g\,Q_g^{-1}$.

However, it is easily seen that for some solutions $\omega_p$ some
of these parameters may be negative. A simple argument based on
elementary inequalities gives the following result.

 \begin{proposition}
 \label{prop1}
Problem \ref{pro3} admits solutions with a Lead-Lag network with
complex poles/zeros if and only if a solution $\omega_p$ of
(\ref{const}) exists such that $\Phi_1$, $\Phi_2$, $\Psi_1$ and
$\Psi_2$ all have the same sign, and
 \begin{itemize}
\item are all positive if $\omega_p<\omega_g$;
\item are all negative if $\omega_p>\omega_g$.
\end{itemize}
Moreover, Problem \ref{pro3} admits solutions with a Lead-Lag
network with real poles/zeros if and only if a solution $\omega_p$
of (\ref{const}) exists such that
 \beann
\max \{ \Phi_1\cdot \Phi_2, \,\Psi_1\cdot \Psi_2 \} <  \dfrac{\omega_g\,\omega_p\,(\omega_g^2-\omega_p^2)}{4}.
\eeann
\end{proposition}
\proof From the expressions of $\zeta_1$, $\zeta_2$ and $\omega_n$,
we see that we must have $\Phi_1$, $\Phi_2$, $\Psi_1$ and $\Psi_2$
all positive when $\omega_p<\omega_g$ and all negative when
$\omega_p>\omega_g$. The second statement follows by imposing
$\zeta_1>1$ and $\zeta_2>1$.\endproof

\subsection{Design examples using Phase-Correction Networks}
\label{ex}
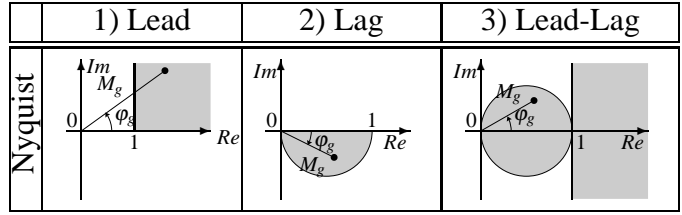
\begin{figure}[tbp]
 \centering
 \begin{tabular}{|c|c|c|c|} \hline
 & 1) Lead & 2) Lag & 3) Lead-Lag \\ \hline\hline
 \rput{90}(0,1){Nyquist} &
 \setlength{\unitlength}{1mm}
 \SpecialCoor
 \psset{unit=\unitlength}
 \psset{arrowlength=.8}
 \psset{arrowinset=.1}
    %%%%%%%%%%%%%%%%
    %%%%%%%%%%%%%%%%
 \begin{picture}(21,20)(-2,-9)
 %\put(10,11){\makebox(0,0)[b]{1) Lead}}
 \pspolygon*[linecolor=verylightgray](7,9)(7,0)(17,0)(17,9)
 \put(-2,0){\vector(1,0){19}}
 \put(0,-9){\vector(0,1){18.5}}
 \put(7,0){\line(0,1){9}}
 \psarc[linewidth=0.3pt]{->}(0,0){4}{0}{38}
 \rput(0,0){\rnode{O}{}}
 \rput(11,8){\rnode{AS}{}}
 \ncline[linewidth=0.3pt]{-*}{O}{AS}\Aput[0.7pt]{\scriptsize $M_g$}
 \put(-0.5,0.5){\makebox(0,0)[br]{\scriptsize $0$}}
 \put(7,-0.5){\makebox(0,0)[t]{\scriptsize $1$}}
 \put(4.35, 0.65){\makebox(0,0)[bl]{\scriptsize $ \varphi_g$}
 %\put(12, 3){\makebox(0,0){\duno}}
 %\put(10, 7.25){\makebox(0,0)[tl]{$M$}}
 \put(-0.5,8){\makebox(0,0)[r]{\scriptsize  $Im$}}
 \put(15,-0.5){\makebox(0,0)[t]{\scriptsize  $Re$}} }
 \end{picture}
 %%%%%%%%%%%%%%%%
 &
 %%%%%%%%%%%%%%%%
 \setlength{\unitlength}{1mm}
 \SpecialCoor
 \psset{unit=\unitlength}
 \psset{arrowlength=.8}
 \psset{arrowinset=.1}
 \begin{picture}(21,20)(-2,-9)
 %\put(10,11){\makebox(0,0)[b]{2) Lag}}
 \psarc*[linecolor=verylightgray](6,0){6}{180}{360}
 \psarc[linewidth=0.3pt](6,0){6}{180}{360}
 \put(-2,0){\vector(1,0){19}}
 \put(0,-9){\vector(0,1){18.5}}
 \psarc[linewidth=0.3pt]{<-}(0,0){4}{-28}{0}
 \rput(0,0){\rnode{O}{}}
 \rput(7,-3.5){\rnode{A}{}}
 \rput(9,-5.196){\rnode{AP}{}}
 \ncline[linewidth=0.3pt]{-*}{O}{A}\Bput[1pt]{\scriptsize $\vspace{-30mm} \hspace{10mm}M_g$}
 \put(-0.5,0.5){\makebox(0,0)[br]{\scriptsize $0$}}
 \put(12,0.5){\makebox(0,0)[b]{\scriptsize $1$}}
 \put(4.35,-0.65){\makebox(0,0)[tl]{\scriptsize $\varphi_g$}}
 %\put(9,-3){\makebox(0,0)[b]{\ddue}}
 { \small
 \put(-0.5,8){\makebox(0,0)[r]{\scriptsize  $Im$}}
 \put(15,-0.5){\makebox(0,0)[t]{\scriptsize  $Re$} }}
 \end{picture}
 %%%%%%%%%%%%%%%%
 &
 %%%%%%%%%%%%%%%%
 \setlength{\unitlength}{1mm}
 \SpecialCoor
 \psset{unit=\unitlength}
 \psset{arrowlength=.8}
 \psset{arrowinset=.1}
 \begin{picture}(26,20)(-2,-9)
 %\put(13,11){\makebox(0,0)[b]{3) Lead-Lag}}
 \psarc*[linecolor=verylightgray](6,0){6}{0}{360}
 \psarc[linewidth=0.3pt](6,0){6}{0}{360}
 \pspolygon*[linecolor=verylightgray](12,9)(12,-9)(22,-9)(22,9)
 \put(-2,0){\vector(1,0){24}}
 \put(0,-9){\vector(0,1){18.5}}
 \put(12,-9){\line(0,1){18}}
 \psarc[linewidth=0.3pt]{->}(0,0){4}{0}{29}
 \rput(0,0){\rnode{O}{}}
 \rput(7, 4.041){\rnode{A}{}}
 \rput(16,8){\rnode{AS}{}}
 %\ncline{-*}{O}{AS}\Bput[1pt]{}
 \ncline[linewidth=0.3pt]{-*}{O}{A}\Aput[0.7pt]{\scriptsize  $\hspace{10mm} M_g$}
 \put(-0.5,0.5){\makebox(0,0)[br]{\scriptsize $0$}}
 \put(12.5,-0.5){\makebox(0,0)[tl]{\scriptsize $1$}}
 \put(4.35, 0.65){\makebox(0,0)[bl]{\scriptsize  $\varphi_g$}}
 %\put(8,-1){\makebox(0,0)[lt]{$D_{3b}$}}
 %\put(8, 1){\makebox(0,0)[lb]{$D_{3c}$}}
 %\put(18, 6){\makebox(0,0)[lb]{$D_{3a}$}}
 %\put(18,-6){\makebox(0,0)[lt]{$D_{3d}$}}
 % \put(15, 7.25){\makebox(0,0)[tl]{$M$}}
 { \small
 \put(-0.5,8){\makebox(0,0)[r]{\scriptsize $Im$}}
 \put(20,-0.5){\makebox(0,0)[t]{\scriptsize $Re$}} }
 \end{picture} \\ \hline
 \end{tabular}
  \caption{\label{nichols} Graphical representation on the
Nyquist plane of $M_g$ and $\varphi_g$ for an admissible solution of
Problems \ref{pro1} and \ref{pro2} using Lead, Lag and
Lead-Lag networks.}
 \end{figure}

In this section our aim is to show how the simple method outlined in
the previous sections can be easily employed to solve a range of
problems that can be fruitfully used as written exercises of a
basic Control course.
First, we notice that when the aim is to assign the gain crossover frequency and the phase margin, we can complement the results in Table \ref{table1} with the considerations on the feasibility of the networks presented in the previous section, see Table \ref{table2}.
\begin{table}[!hbp]
\begin{center}
\vspace{.2cm}
\begin{tabular}{|@{\;}c@{\;}||@{}c@{}|@{}c @{}|}
\hline
 &   $\phantom{\big|}M_g>1$   &  $M_g<1$  \\ \hline \hline
 $\varphi_g \in \left( -\frac{\pi}{2},0\right)$ &    \begin{minipage}{3.5cm}\centering $\phantom{\big|}M_g\cos\varphi_g>1$  \\ Lead-Lag \!\!$\phantom{\big|}(\zeta_1 \!>\! \zeta_2$)\end{minipage}      &
                     \begin{minipage}{3.5cm}\centering  $\phantom{\big|}\cos\varphi_g >M_g$ \\ Lag\! or\! Lead-Lag \!\!$\phantom{\big|}(\zeta_2 \!>\! \zeta_1$) \end{minipage}    \\ \hline
 $\varphi_g \in \left( 0,\frac{\pi}{2}\right)$ &    \begin{minipage}{3.5cm}\centering $\phantom{\big|}M_g\cos\varphi_g>1$  \\ Lead or Lead-Lag \!\!$\phantom{\big|}(\zeta_1 \!>\! \zeta_2$)  \end{minipage}      &
                     \begin{minipage}{3.5cm}\centering  $\phantom{\big|}\cos\varphi_g >M_g$ \\ Lead-Lag \!\!$\phantom{\big|}(\zeta_2 \!>\! \zeta_1$)  \end{minipage}    \\ \hline
\end{tabular}
\end{center}
\caption{$\qquad$}
\label{table2}
\ \vspace{-1cm}
\end{table}
A graphical
representation -- that complements the use of Table \ref{table2} --
of the points $M_g\,e^{j \varphi_g}$ of the Nyquist plane for which
the problem admits solutions with a Lead, Lag or Lead-Lag network is
given in Figure \ref{nichols}.

%%%%%%%%%%%%%%%%%%%%%%%%%%%%%%%%%%%%%%%%%%%%%%%%%%%%%%%%%%%%%%%%
 %%%%%%%%%%%%%%%%%%%%%%%%%%%%%%%%%%%%%%%%%%%%%%%%%%%%%%%%%%%%%%%%

This figure provides a useful mean to gain insight into the design procedure presented here. The desired gain crossover frequency $\omega_g$ defines a point $A$ on the Nyquist plot of the plant, i.e., $A=\bar{G}(j \omega_g)$. The specification on the phase margin defines a point $B$ on the unit circle that the loop gain has to cross at exactly the same frequency, i.e., $B=e^{j\,(\pi+\textrm{PM})}=L(j\,\omega_g)$. As such, the design reduces to finding the compensator structure such that $C(j \omega_g)\,A=B$. Loosely speaking, we may say that the network brings point $A$ into point $B$ at the frequency $\omega_g$. The solution of this problem is exactly the one given by the inversion formulae. The feasibility of each type of network imposes a constraint on $M_g$ and $\varphi_g$, i.e., on the position that point $A$ must have with respect to point $B$ in order for a network with positive parameters to exist. These feasibility constraints are represented graphically by the shaded regions in Figure \ref{nichols}.

\begin{algorithm}[h!]
   \caption{Solution of Question 1 in MATLAB$^{\textrm{\tiny{\textregistered}}}$}
     \begin{algorithmic}[1]
     \label{Alg}
     \STATE {\small \begin{verbatim}s=tf('s'); \end{verbatim}}
    \STATE {\small \begin{verbatim}G=0.5*(s+10)/(s*(s^2+2*s+10));\end{verbatim}}
    \STATE {\small \begin{verbatim}wg=3;\end{verbatim}}
    \STATE {\small \begin{verbatim}PM=pi/4;\end{verbatim}}
    \STATE {\small \begin{verbatim}C=evalfr(G,j*wg);\end{verbatim}}
    \STATE {\small \begin{verbatim}M=1/abs(C);\end{verbatim}}
    \STATE {\small \begin{verbatim}phi=PM-(pi+angle(C));\end{verbatim}}
    \STATE {\small \begin{verbatim}if (sin(phi)<0)|(cos(phi)<0)|M<1/cos(phi),\end{verbatim}}
        \STATE {\small \begin{verbatim}  disp('No solutions with a Lead network');\end{verbatim}}
        \STATE {\small \begin{verbatim}  return\end{verbatim}}
    \STATE {\small \begin{verbatim}end\end{verbatim}}
    \STATE {\small \begin{verbatim}alpha=(M*cos(phi)-1)/(M*(M-cos(phi)));\end{verbatim}}
    \STATE {\small \begin{verbatim}tau=(M-cos(phi))/(wg*sin(phi));\end{verbatim}}
     \end{algorithmic}
 \end{algorithm}

\begin{figure}[htb]
\centering
\psfrag{R}[]{\small $R(s)$}
\psfrag{E}[]{\small $E(s)$}
\psfrag{G}[]{\small $Y(s)$}
\psfrag{I}[]{\tiny $+$}
\psfrag{J}[]{\tiny $-$}
\psfrag{U}[]{\small $U(s)$}
\psfrag{H}[]{\small $\ssp \,\, H(s) \ssp$}
\psfrag{B}[]{\small $\ssp \,\, C(s) \ssp$}
\psfrag{A}[]{\small $\ssp \,\, \dfrac{s+10}{\ssp s\,(s^2+2\,s+10)\ssp} \ssp$}
\includegraphics[width=3.6in]{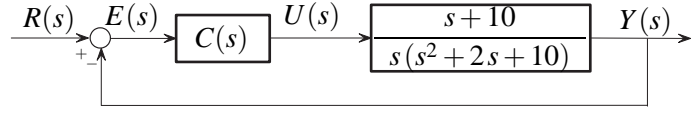}
\caption{Unity feedback control scheme.}
\label{fig3}
\end{figure}

%The range of problems/questions/exercises that can be formulated goes far beyond those that can be
Consider the control scheme in Figure \ref{fig3}.
%%%%%%%%%%%%%%%%%%%%%%%%%%%%%%%%%%%%
%%%%%%%   FIGURE   %%%%%%%%%%%%%%%%%
%%%%%%%%%%%%%%%%%%%%%%%%%%%%%%%%%%%%
%

{\bf Question 1}. Design a phase-correction network that satisfies the following static and dynamic specifications:

\begin{itemize}
\item velocity constant equal to $0.5$;
\item phase margin equal to $45^\circ$;
\item gain crossover frequency equal to $3$ rad/sec.
\end{itemize}
Find also the range of phase margins that are achievable at the crossover frequency $3$ rad/sec with this phase-correction network.
Also, determine the range of phase margins that at this gain crossover frequency ensures closed-loop stability.\\
\begin{figure}[htb]
\centering
 \psfrag{Imag}[tl][tl][0.6]{$$}
 \psfrag{Real}[tl][tl][0.6]{$$}
 \psfrag{A}[tl][tl][1.0]{$A$}
 \psfrag{B}[tl][tl][1.0]{$B$}
 \psfrag{Diagramma di Nyquist}[tl][tl][1.0]{$$}
 %%%%%%
 \psfrag{G}[tr][tr]{$\bar{G}(j\omega)$}
 \psfrag{L}[br][br]{$L(j\omega)$}
 \psfrag{I}[r][r][0.7]{Im}
 \psfrag{R}[r][r][0.7]{Re}
 %%>>%%  nome  = gs_Lorenzo_Lead
 %%>>%%  NY.Show=Si
 %%>>%%  gs=0.5*(s+10)/(s*(s^2+2*s+10))
 %%>>%%  cs=1
 %%>>%%  cs=(1+2.6317*s)/(1+0.6817*s)
 %%>>%%  NY.Plot_LineStyle='k'
 %%>>%%  NY.Plot_LineStyle='-r'
 %%>>%%  NY.Show_Cs=Si
 %%>>%%  NY.Axes_MyGrid=Si
 %%>>%%  NY.Cs_Tipo='Lead'
 %%>>%%  NY.Cs_Fill_Color=0.85
 %%>>%%  NY.Cs_WA=[3]
 %%>>%%  NY.Cs_B=[1 180+45]
 %%>>%%  NY.Cs_Show_A=Si
 %%>>%%  NY.Cs_A_Posizione='lm'
 %%>>%%  NY.Cs_Show_B=Si
 %%>>%%  NY.Cs_B_Posizione='rt'
 %%>>%%  NY.Cs_Show_Freccia_AB=Si
 %%>>%%  NY.Show_W_Ticks=Si
 %%>>%%  NY.W_Ticks=[2 2.7 3 3.3 ]
 %%>>%%  NY.W_Ticks_FontSize=9
 %%>>%%  NY.W_Ticks_Side=1
 %%>>%%  NY.W_Ticks_Side=0
 %%>>%%  NY.Show_Axes=No
 %%>>%%  NY.Axes_Re=[-1.4 0.2]
 %%>>%%  NY.Axes_Im=[0.2]
 %%>>%%  NY.Axes_Clip=Si
 %%>>%%  NY.Frecce=[3.8 0.8]
 %%>>%%  NY.Frecce=[3.8 2.9]
 %%>>%%  device= -depsc
 %%>>%%  NY.Esegui= text(-0.1005,   -0.6860,''G'')
 %%>>%%  NY.Esegui= text(-0.6793,   -0.8439,''L'')
 %%>>%%  NY.Esegui= text(-0.0267,    0.1596,''I'')
 %%>>%%  NY.Esegui= text( 0.1650,   -0.0333,''R'')
  %%%%%%%%%%%%%%%%
\includegraphics[width=7.3cm]{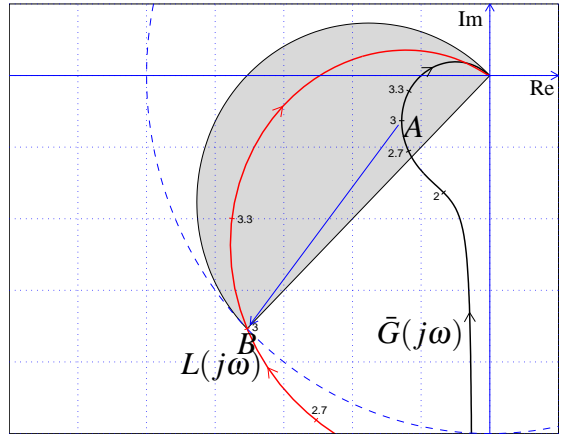}
\caption{Graphical representation on the Nyquist of solution of
Question $1$.} \label{NY_gs_Lorenzo_Lead}
\end{figure}
The DC gain of the phase-correction network $K$ must be selected so
as to satisfy the specification on the velocity constant:
\[
K_v=\lim_{s \to 0} s\,C(s)\,G(s)= \lim_{s \to 0} s\,K\,G(s)=K,
\]
so that $K=0.5$. The gain $K$ is now considered to be part of the
plant, i.e., we define $\bar{G}(s)=K\,G(s)$. In order to select the
right compensation structure, we compute $M_g$ and $\varphi_g$:
\beann
M_g \ns&\ns = \ns&\ns \dfrac{1}{|\bar{G}(3\,j)|} = 6 \sqrt{\dfrac{37}{109}}\simeq 3.4957 >1, \\
\varphi_g \ns&\ns = \ns&\ns \textrm{PM}-(\pi+\textrm{arc} \bar{G}(3\,j)) \\
\ns&\ns = \ns&\ns \frac{7}{4}\pi-\arctan (3/10) + \arctan 6 \simeq 18.84^\circ.
\eeann

As such, using Table \ref{table1} we see that a Lead network may be
used. Since the conditions (\ref{conditionsLead}) are both
satisfied, we expect the problem to be solvable. A simple
computation, that can even be carried out in closed-form with pen
and paper, shows that

\beann
\alpha\ns&\ns = \ns&\ns \dfrac{3 \cdot 85 \cdot \sqrt{2}-109}{36\cdot 37 -3 \cdot 85\sqrt{2}}\simeq 0.2590,\\
\tau \ns&\ns = \ns&\ns \dfrac{12 \cdot 37-85\sqrt{2}}{3\cdot 29
\sqrt{2}} \simeq 2.6317\,\textrm{sec}. \eeann The corresponding
MATLAB$^{\textrm{\tiny{\textregistered}}}$ instructions are shown in
Algorithm 1, and the required compensator that satisfies all the
specifications is given by \beann C_{\rm Lead}(s)=0.5\,
\dfrac{1+2.6317\,s}{1+0.6817\,s}. \eeann
A graphical plot on the Nyquist plane of the frequency response
$\overline{G}(j\omega)$ is shown with the black line in
Fig.~\ref{NY_gs_Lorenzo_Lead}, where $A$ denotes the point of
$\overline{G}(j\omega)$ at frequency $\omega_g=3$ rad/sec. The
compensator $C_{\rm Lead}(s)$ has been designed such that
$L(j\omega)$ shown with red line passes through point
$B=e^{j\,(\textrm{$PM+\pi$})}$ at frequency $\omega_g$. Intuitively,
we can say that the point $A$ is brought to point $B$ by
multiplication with the compensator frequency response at
$\omega=\omega_g$. The gray area in Fig.~\ref{NY_gs_Lorenzo_Lead}
denotes the set of all the points that can be brought to the desired
point $B$ using a Lead network.

The smallest phase margin achievable with a Lead network at the gain crossover frequency $\omega_g=3$ rad/sec is
\[
\textrm{PM}_{\rm min}=\pi+\textrm{arg} \bar{G}(j \omega_g)=\dfrac{\pi}{2}+\arctan \dfrac{3}{10}-\arctan 6 \simeq  26.1616^\circ,
\]
and the largest phase margin is
\beann
\textrm{PM}_{\rm max} \ns&\ns = \ns&\ns \pi+\textrm{arg} \bar{G}(j \omega_g)+\arccos (|\bar{G}(j \omega_g)|)\\
\ns&\ns = \ns&\ns \dfrac{\pi}{2}+\arctan \dfrac{3}{10}-\arctan 6+\arccos \dfrac{1}{6} \sqrt{\dfrac{109}{37}} \simeq  99.54^\circ.
\eeann

{\bf Question 2}. Design a phase-correction network that satisfies the following specifications:

\begin{itemize}
\item velocity error equal to $0.1$;
\item phase margin equal to $60^\circ$;
\item gain crossover frequency equal to $1$ rad/sec.
\end{itemize}
Find also the range of phase margins that are achievable at this crossover frequency with this phase-correction network.\\
\begin{figure}[htb]
\centering
 \psfrag{Imag}[tl][tl][0.6]{$$}
 \psfrag{Real}[tl][tl][0.6]{$$}
 \psfrag{A}[tl][tl][1.0]{\hspace{1mm}$A$}
 \psfrag{B}[tl][tl][1.0]{\hspace{-3mm}$B$}
 \psfrag{Diagramma di Nyquist}[tl][tl][1.0]{$$}
%%%%%%%%%%%%
 \psfrag{G}[tr][tr]{$\bar{G}(j\omega)$}
 \psfrag{L}[br][br]{$L(j\omega)$}
 \psfrag{I}[r][r][0.7]{Im}
 \psfrag{R}[r][r][0.7]{Re}
 %%%%%%%%%%%%%%%%
 %%>>%%  nome  = gs_Lorenzo_Lag
 %%>>%%  NY.Show=No
 %%>>%%  gs=10*(s+10)/(s*(s^2+2*s+10))
 %%>>%%  cs=1
 %%>>%%  cs=(1+2.102*s)/(1+25.36*s)
 %%>>%%  NY.Plot_LineStyle='k'
 %%>>%%  NY.Plot_LineStyle='r'
 %%>>%%  NY.Show_Cs=Si
 %%>>%%  NY.Axes_MyGrid=Si
 %%>>%%  NY.Cs_Tipo='Lag'
 %%>>%%  NY.Cs_Fill_Color=0.85
 %%>>%%  NY.Cs_WA=[1]
 %%>>%%  NY.Cs_B=[1 180+60]
 %%>>%%  NY.Cs_Show_A=Si
 %%>>%%  NY.Cs_A_Posizione='lm'
 %%>>%%  NY.Cs_Show_B=Si
 %%>>%%  NY.Cs_B_Posizione='rm'
 %%>>%%  NY.Cs_Show_Freccia_AB=Si
 %%>>%%  NY.Cs_Freccia_AB_Spazio=0.03
 %%>>%%  NY.Show_W_Ticks=Si
 %%>>%%  NY.W_Ticks=[1 1.7 2.7 3.3]
 %%>>%%  NY.W_Ticks=[0.2 0.33 0.56 1 ]
 %%>>%%  NY.W_Ticks_FontSize=9
 %%>>%%  NY.W_Ticks_Side=1
 %%>>%%  NY.W_Ticks_Side=0
 %%>>%%  NY.Show_Axes=No
 %%>>%%  NY.Axes_Re=[-12 6]
 %%>>%%  NY.Axes_Re_Tick=[-12:2:6]
 %%>>%%  NY.Axes_Im=[2.0]
 %%>>%%  NY.Axes_Im_Tick=[-12:2:6]
 %%>>%%  NY.Axes_Clip=Si
 %%>>%%  NY.Frecce=[3.1 1.4]
 %%>>%%  NY.Frecce=[0.22]
 %%>>%%  device= -depsc
 %%>>%%  NY.Esegui= text( -1.9839,   -8.8750,''G'')
 %%>>%%  NY.Esegui= text( -8.0806,   -4.7303,''L'')
 %%>>%%  NY.Esegui= text( -0.3249,    1.5855,''I'')
 %%>>%%  NY.Esegui= text(  5.1912,   -0.3882,''R'')
  %%%%%%%%%%%%%%%%
\includegraphics[width=7.3cm]{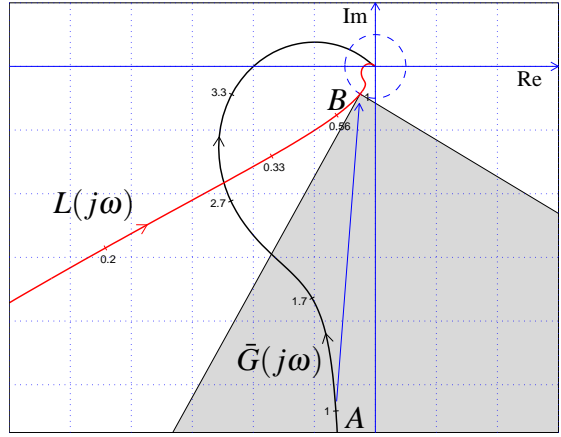}
\caption{Design of a Lag network on the Nyquist plane to meet the
specifications of Question $2$.} \label{NY_gs_Lorenzo_Lag}
\end{figure}

Since the velocity error is equal to $e_v=1/K_v$ and $K_v=K$ as shown in Question 1, it is found that $K=10$.
We define $\bar{G}(s)=K\,G(s)$. In order to select the right compensation structure, we compute $M_g$ and $\varphi_g$:
\beann
M_g \ns&\ns = \ns&\ns = \dfrac{1}{10} \sqrt{\dfrac{85}{101}}\simeq 0.0917< 1, \\
\varphi_g \ns&\ns = \ns&\ns -\frac{\pi}{6}-\arctan (1/10) + \arctan
\dfrac{2}{9} \simeq -23.18^\circ. \eeann Therefore, we can use a Lag
network. Since $-\arccos(M_g)<\varphi_g<0$, a Lag network solving
the problem exists, and is characterised by the parameters
$\alpha=0.0829$ and $\tau=25.3559$ sec by simply replacing $M_g$ and
$\varphi_g$ thus found into the Inversion Formulae
(\ref{inversionLag}) for the Lag network.

The smallest phase margin achievable with a Lag network at the gain crossover frequency $\omega_g=1$ rad/sec is
\[
\textrm{PM}_{\rm min}=\pi+\textrm{arg} \bar{G}(j \omega_g)-\arccos \dfrac{1}{|\bar{G}(j \omega_g)|} \simeq  -1.55^\circ,
\]
and the largest phase margin is
\[
\textrm{PM}_{\rm max}=\pi+\textrm{arg} \bar{G}(j \omega_g) \simeq  83.18^\circ.
\]
Intuitively, the designed Lag network brings the point $A=\bar{G}(j
\omega_g)$ to the desired point $B=e^{j\,240^\circ}$ as shown in
Fig.~\ref{NY_gs_Lorenzo_Lag}. The gray area denotes the set of
points that can be brought to $B$ by a Lag network. If $A=\bar{G}(j
\omega_g)$ is not within this area, the problem cannot be solved
with this type of network.

{\bf Question 3}. Let $K=10$. Find the interval of gain margins achieved using a Lag network at the phase crossover frequency
$\omega_p=4$ rad/sec that guarantee asymptotic stability of
the closed loop.

A simple computation shows that
\beann
M_p \ns&\ns = \ns&\ns \dfrac{2}{\textrm{GM}\sqrt{29}}\\
\varphi_p \ns&\ns = \ns&\ns  -\dfrac{\pi}{2}+\arctan
\dfrac{2}{5}+\arctan \dfrac{4}{3} \eeann from which it follows that
\beann \alpha=
\dfrac{52-\frac{20}{\textrm{GM}}}{145\,\textrm{GM}-52}, \qquad
\tau=\dfrac{145\,\textrm{GM}-52}{56}. \eeann It follows that \beann
C(s)=\dfrac{56+\left(52-\frac{20}{\textrm{GM}}\right)s}{56+(145\,\textrm{GM}-52)\,s}.
\eeann The characteristic polynomial is \beann
&&\hspace{-3mm} (145\,\textrm{GM}-52)\,s^4+(290\,\textrm{GM}-48)\,s^3 \\
&&\hspace{-3mm}
+(112+1450\textrm{GM}-\frac{200}{\textrm{GM}})\,s^2+(6320-\frac{2000}{\textrm{GM}})\,s+5600=0.
\eeann The asymptotic stability of the closed loop can at this point
be studied using the Routh criterion on this polynomial. Such study
will lead to a set of intervals for $\textrm{GM}$ that guarantee
asymptotic stability of the closed-loop.\\[-2mm]

{\bf Question 4}.
Design a Lead-Lag network that meets the following specifications:
\begin{itemize}
\item velocity constant equal to $0.1$;
\item phase margin equal to $45^\circ$;
\item gain margin equal to $3$;
\item gain crossover frequency equal to $1$ rad/sec.
\end{itemize}
\begin{figure}[htb]
\centering
 \psfrag{Im}[tl][tl][0.8]{$Im$}
 \psfrag{Re}[tl][tl][0.8]{$Re$}
 \psfrag{G}[tl][tl][1.0]{\hspace{1mm}$G(j\omega)$}
 \psfrag{L}[tl][tl][1.0]{$L(j\omega)$}
 \psfrag{w1}[tl][tl][0.5]{\hspace{2mm}$2.3$}
 \psfrag{w2}[tl][tl][0.5]{\hspace{0mm}$2.3$}
 \psfrag{1.0}[tl][tl][0.5]{\hspace{1mm}$1$}
 \psfrag{A}[tr][tr][0.8]{\hspace{-2mm}$A$}
 \psfrag{B}[tl][tl][0.8]{\hspace{2mm}$B$}
 \psfrag{C}[tl][tl][0.8]{\hspace{-2mm}$C$}
 \psfrag{D}[tl][tl][0.8]{\hspace{-1mm}$D$}
 \psfrag{Diagramma di Nyquist}[tl][tl][1.0]{$$}
\includegraphics[width=7.3cm]{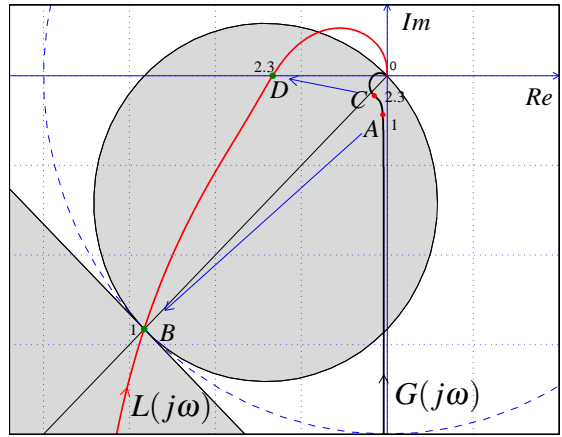}
\caption{Design of Lead-Lag network on the Nyquist plane to meet
the specifications of Question $4$.} \label{Lead_Lag_quest4}
\end{figure}
First, notice that only a Lead-Lag network can simultaneously meet all the specifications.
The DC gain of the Lead-Lag network $K$ must be selected so as to satisfy the specification on the velocity constant:
\[
K_v=\lim_{s \to 0} s\,C_{\rm LL}(s)\,G(s)= \lim_{s \to 0} s\,K\,G(s)=K,
\]
so that $K=0.1$. Now, we consider this gain to be part of the plant, and define $\bar{G}(s)=K\,G(s)$. The frequency response of the plant at $\omega=\omega_g$ is
\[
G(j \omega_g)= \frac{10+j}{j\,(2\,j+9)},
\]
which leads to \beann
M_g \ns&\ns = \ns&\ns \dfrac{1}{K} \sqrt{\dfrac{85}{101}}, \\
\varphi_g \ns&\ns = \ns&\ns \dfrac{\pi}{4} - \pi - (\arctan \dfrac{1}{10}-\frac{\pi}{2}-\arctan \frac{2}{9}) \\
\ns&\ns = \ns&\ns \dfrac{3}{4}\,\pi-\arctan \dfrac{1}{10}+\arctan
\frac{2}{9}. \eeann A simple goniometric calculation shows that
\beann \cos \varphi_g = \dfrac{103}{2} \sqrt{\dfrac{2}{85\cdot 101}}
\eeann which leads to \beann \gamma = \dfrac{M_g-\cos
\varphi_g}{\cos
\varphi_g-M_g^{-1}}=\dfrac{170-103\,\sqrt{2}\,K}{103\,\sqrt{2}\,K-202\,K^2}.
\eeann We express $M_p$ and $\varphi_p$ as a function of the phase
crossover frequency $\omega_p$, which is still unknown: \beann
M_p \ns&\ns = \ns&\ns \dfrac{\omega_p\,\sqrt{4\,\omega_p^2+(10-\omega_p^2)^2}}{GM \cdot K \sqrt{\omega_p^2+100}} \\
\cos \varphi_p\ns&\ns = \ns&\ns \dfrac{\omega_p\,(\omega_p^2+10)}{\sqrt{(100+\omega_p^2)((10-\omega_p^2)^2+4\,\omega_p^2)}}
\eeann
Plugging these expressions into (\ref{const}) yields the polynomial equation
 \beann
\omega^6-(16+H+\gamma  H)\omega^4+10\,(10-H-\gamma
H)\,\omega^2+100\,\gamma\,H^2=0
 \eeann
where $H=\textrm{GM} \cdot K$. This biquadratic equation has two
positive solutions: $\omega_p^\prime=3.9591$ and $\omega_p^{\prime
\prime}=2.3686$. Using the first solution, we obtain positive values
for $\Phi_1$, $\Phi_2$, $\Psi_1$ and $\Psi_2$. Therefore, the
condition of Proposition \ref{prop1} are not satisfied. If we use
$\omega_p=\omega_p^{\prime \prime}=2.3686$, those values are all
negative, and therefore they satisfy the condition of Proposition
\ref{prop1}. The corresponding values of the parameters of the
Lead-Lag network are $\zeta_1=20.7474$, $\zeta_2=1.6747$ and
$\omega_n= 0.2980$. Since $\zeta_1$ and $\zeta_2$ are both greater
than $1$, the solution can be also given in terms of a Lead-Lag
network with real poles/zeros, with $\alpha=0.0728$, $\tau_1=1.1120$
sec and $\tau_2=139.1776$ sec. \\

Intuitively, point $A=\bar G(j \omega_g)$ is brought to $B=e^{j\cdot
225^\circ}=L(j\omega_g)$ and point $C=\bar G(j\omega_p^{\prime
\prime})$ is brought to $D=\frac{e^{j\,180^\circ}}{\textrm{GM}}$.
The gray area in Fig.~\ref{Lead_Lag_quest4} represents the set of
all points that can be brought to  $B$ using a Lead-Lag network.

This example has shown the effectiveness of this design method from
another perspective. In fact, even in the case of a plant with a
reasonably rich dynamical structure, the design problem is found to
admit a closed-form solution. Indeed, the $6{\rm th}$ degree
polynomial equation in $\omega_p$ is biquadratic, and is therefore
solvable in finite terms. This leads to a ``mathematically exact''
solution of this design problem. To the best of the authors'
knowledge, this is the first time an exact explicit solution for
this problem is given for a Lead-Lag network.

\section{PID controllers}
Consider the example in Section \ref{ex} where system $G(s)$ is of
type $1$. Lead and Lag networks are not suitable compensators for a
design problem involving steady-state specifications that require
zero velocity error (i.e., that the resulting control system tracks
a ramp reference signal with zero steady-state error). In this case,
a compensator that meets the desired steady-state specification is a
PID controller, because of the presence of a pole at the origin in
its transfer function. If the steady-state specification simply
consists in achieving zero velocity error, the use alone of a
compensator with a pole at the origin is sufficient to guarantee
that this requirement is satisfied. In this case, the steady-state
specification does not lead to constraints on the parameters of the
controller. However, there are steady-state requirements that lead
to such constraints. For example, if the controller is required to
guarantee that the acceleration error be no greater than, say,
$0.2$, we obtain a constraint on the ratio $K_p/T_i$, which is known
also as {\em integration constant} of the PID controller. In fact,
 \beann
e_a= \lim_{s \to 0} \dfrac{1}{s^2\,C_{\rm PID}(s) G(s)}=\dfrac{T_i}{K_p} \le 0.2
 \eeann
leads to $K_p/T_i \ge 5$. Hence, in this case the ratio $K_p/T_i$ is
assigned by the steady-state requirements. These two situations must
be addressed separately, because of the significant differences
arising in the design phase. \\

\begin{figure}[tbp]
 \centering
 \begin{tabular}{|c|c|c|c|} \hline
 & 1) PID & 2) PD & 3) PI \\ \hline\hline
 \rput{90}(0,1.5){Nyquist} &
 %\setlength{\unitlength}{1mm}
% \SpecialCoor
% \psset{unit=\unitlength}
% \psset{arrowlength=.8}
% \psset{arrowinset=.1}
 \setlength{\unitlength}{1.0cm}
 \psset{unit=\unitlength}
    %%%%%%%%%%%%%%%%
    %%%%%%%%%%%%%%%%
 \begin{pspicture}(-0.1,-1.5)(2,1.7)
%%%%%%%%%%ASSI%%%%%
 \pscustom[linewidth=0pt,fillstyle=solid,fillcolor=verylightgray]{
    \psline[linewidth=0.5pt](1.8,1.4)(0,1.4)
    \psline[linewidth=0.5pt](0,1.4)(0,-1.4)
    \psline[linewidth=0.5pt](0,-1.4)(1.8,-1.4)
    \psline[linewidth=0.5pt](1.8,-1.4)(1.8,1.4)
 }
 \psline[linewidth=0.5pt]{->}(-0.3,0)(2,0)
 \psline[linewidth=0.5pt]{->}(0,-1.5)(0,1.5)
 \psline[linewidth=0.5pt]{->}(0,0)(1,0.7)
 \psline[linewidth=0.8pt,linecolor=red](1,-1.4)(1,1.4)
 \psline[linewidth=0.8pt,linestyle=dashed](0.4,-1.4)(0.4,1.4)
 \psline[linewidth=0.8pt,linestyle=dashed](1.6,-1.4)(1.6,1.4)
 \psline[linewidth=0.5pt]{<->}(0.2,-0.2)(1.7,-0.2)
 \psline[linewidth=0.5pt]{<->}(1.1,0.5)(1.1,1)
 \rput[r](-0.1,1.4){\scriptsize $Im$}
 \rput[t](1.9,-0.05){\scriptsize $Re$}
 \rput[rt](-0.1,-0.05){\scriptsize $0$}
 \rput[l](1.05,-0.4){\scriptsize $K_P$}
 \rput[l](1.05,-1.0){\scriptsize $r$}
 \rput[l](1.15,0.5){\scriptsize $\frac{1}{T_i}$}
 \rput[l](1.15,0.9){\scriptsize $T_d$}
 \rput[l](0.55,0.7){\scriptsize $\hspace{-0.8mm}M_g$}
 \rput[l](0.45,0.15){\scriptsize $\varphi_g$}
\end{pspicture}
 %%%%%%%%%%%%%%%%
 &
 \setlength{\unitlength}{1.0cm}
 \psset{unit=\unitlength}
    %%%%%%%%%%%%%%%%
    %%%%%%%%%%%%%%%%
 \begin{pspicture}(-0.1,-1.5)(2,1.7)
%%%%%%%%%%ASSI%%%%%
 \pscustom[linewidth=0pt,fillstyle=solid,fillcolor=verylightgray]{
    \psline[linewidth=0.5pt](1.8,1.4)(0,1.4)
    \psline[linewidth=0.5pt](0,1.4)(0,0)
    \psline[linewidth=0.5pt](0,0)(1.8,0)
    \psline[linewidth=0.5pt](1.8,0)(1.8,1.4)
 }
 \psline[linewidth=0.5pt]{->}(-0.3,0)(2,0)
 \psline[linewidth=0.5pt]{->}(0,-1.5)(0,1.5)
 \psline[linewidth=0.5pt]{->}(0,0)(1,0.7)
 \psline[linewidth=0.8pt,linecolor=red](1,0)(1,1.4)
 \psline[linewidth=0.8pt,linestyle=dashed](0.4,0)(0.4,1.4)
 \psline[linewidth=0.8pt,linestyle=dashed](1.6,0)(1.6,1.4)
 \psline[linewidth=0.5pt]{<->}(0.2,-0.2)(1.7,-0.2)
 \psline[linewidth=0.5pt]{->}(1.1,0.5)(1.1,1)
 \rput[r](-0.1,1.4){\scriptsize $Im$}
 \rput[t](1.9,-0.05){\scriptsize $Re$}
 \rput[rt](-0.1,-0.05){\scriptsize $0$}
 \rput[l](1.05,-0.4){\scriptsize $K_P$}
% \rput[l](1.05,-1.0){\scriptsize $r$}
 \rput[l](1.15,0.85){\scriptsize $T_d$}
 \rput[l](0.55,0.7){\scriptsize $\hspace{-0.8mm}M_g$}
 \rput[l](0.45,0.15){\scriptsize $\varphi_g$}
\end{pspicture}
 %%%%%%%%%%%%%%%%
 &
 \setlength{\unitlength}{1.0cm}
 \psset{unit=\unitlength}
    %%%%%%%%%%%%%%%%
    %%%%%%%%%%%%%%%%
 \begin{pspicture}(-0.1,-1.5)(2,1.7)
%%%%%%%%%%ASSI%%%%%
 \pscustom[linewidth=0pt,fillstyle=solid,fillcolor=verylightgray]{
    \psline[linewidth=0.5pt](1.8,0)(0,0)
    \psline[linewidth=0.5pt](0,0)(0,-1.4)
    \psline[linewidth=0.5pt](0,-1.4)(1.8,-1.4)
    \psline[linewidth=0.5pt](1.8,-1.4)(1.8,-1.4)
 }
 \psline[linewidth=0.5pt]{->}(-0.3,0)(2,0)
 \psline[linewidth=0.5pt]{->}(0,-1.5)(0,1.5)
 \psline[linewidth=0.5pt]{->}(0,0)(1,-0.7)
 \psline[linewidth=0.8pt,linecolor=red](1,-1.4)(1,0)
 \psline[linewidth=0.8pt,linestyle=dashed](0.4,-1.4)(0.4,0)
 \psline[linewidth=0.8pt,linestyle=dashed](1.6,-1.4)(1.6,0)
 \psline[linewidth=0.5pt]{<->}(0.2,0.2)(1.7,0.2)
 \psline[linewidth=0.5pt]{->}(1.1,-0.5)(1.1,-1)
 \rput[r](-0.1,1.4){\scriptsize $Im$}
 \rput[t](1.9,-0.05){\scriptsize $Re$}
 \rput[rt](-0.1,-0.05){\scriptsize $0$}
 \rput[l](1.05,0.4){\scriptsize $K_P$}
 %\rput[l](1.05,-1.0){\scriptsize $r$}
 \rput[l](1.15,-0.8){\scriptsize $\frac{1}{T_i}$}
 \rput[l](0.55,-0.7){\scriptsize $\hspace{-0.8mm}M_g$}
 \rput[l](0.45,-0.17){\scriptsize $\varphi_g$}
\end{pspicture}
 %%%%%%%%%%%%%%%%
   \\\hline
  &$M_g>0$
 &
 $M_g>0$
 &
 $M_g>0$\\

& $\varphi_g\in(-\frac{\pi}{2},\frac{\pi}{2})$
 &
 $\varphi_g\in(0,\frac{\pi}{2})$
 &
 $\varphi_g\in(-\frac{\pi}{2},0)$
 \\ \hline
 \end{tabular}
  \caption{\label{nichols1} Graphical representation on the
Nyqsuit plane of admissible values of $M_g$ and $\varphi_g$ for PID,
PD and PI compensators.}
 \end{figure}
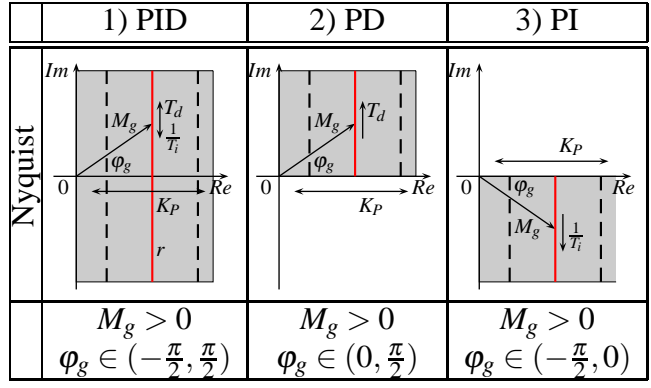

The graphical representation of the frequency response of PID
controllers on the Nyquist plane, usually omitted in undergraduate
textbooks, is helpful to understand the physical meaning of the
regulator synthesis, see Fig.~\ref{nichols1}. The frequency response
of a PID controller is a vertical line passing through point
$(K_p,0)$. Variations of parameters determine the gray area of
admissible values of $M_g$ and $\varphi_g$ useful in the synthesis
procedure of
the compensator. \\

First, we consider the case where the steady-state specifications do
not lead to a constraint on the integral constant of the PID
controller.
%In order to compute the parameters of the PID controller, we write ${G}(j \omega)$ and $C_{\rm PID}(j \omega)$ in polar form as
%\beann
%{G}(j \omega)= |{G}(j \omega)|\,e^{\,j\,\textrm{arg}\,{G}(j \omega)}, \quad   C_{\rm PID}(j \omega)=M(\omega)\,e^{\,j\,\varphi(\omega)}.
%\eeann
%The loop gain frequency response can be written as $L(j \omega)= |{G}(j \omega)|\,M(\omega)\,e^{\,j\,\left(\textrm{arg}\,{G}(j \omega)+\varphi(\omega)\right)}$.
%If the gain crossover frequency $\omega_g$ and the phase margin $\textrm{PM}$ of the loop gain transfer function $L(s)$ are assigned, by (\ref{L1}-\ref{L2}) it is found that $|L(j \omega_g)| = 1$ and $\textrm{PM}  = \pi+\textrm{arg}\,L(j \omega_g)$ must be satisfied. These two equations can be written as
%\begin{enumerate}
%\item $M_g= {1}/{\phantom{\Big|}|{G}(j \omega_g)|\phantom{\Big|}}$,
%\item $\varphi_g=\textrm{PM}-\pi-\textrm{arg}\,{G}(j \omega_g)$,
%\end{enumerate}
%where $M_g\stackrel{\text{\tiny def}}{=}M(\omega_g)$ and $\varphi_g\stackrel{\text{\tiny def}}{=}\varphi(\omega_g)$.

%%%%%%%%%%%%%%%%%%%%%%%%%%%%%%%%%%%%%%%%%%%%%%%%%%%%%%%%%%%%%%%%
 %%%%%%%%%%%%%%%%%%%%%%%%%%%%%%%%%%%%%%%%%%%%%%%%%%%%%%%%%%%%%%%%

In order to find the parameters of the controller such that {\em
i-a)} and {\em ii-a)} are met, equation (\ref{inversion}), that in
the case $\bar{C}(s)=C_{\rm PID}(s)$ becomes \bea \label{eqpar}
M_g\,e^{\,j\,\varphi_g}=K_p\,\dfrac{1+j\,\omega_g\,T_i-\omega_g^2\,T_i\,T_d}{j\,\omega_g\,T_i},
\eea which must be solved in $K_p,T_i,T_d >0$. By equating real and
imaginary parts of both sides of (\ref{eqpar}) we get \bea
&& \omega_g\,M_g\,T_i\,\cos \varphi_g = \omega_g\,K_p\,T_i, \label{eqaa} \\
&& -M_g\,\omega_g\,T_i\,\sin \varphi_g = K_p-K_p\,\omega_g^2\,T_i\,T_d, \label{eqbb}
 \eea
in the three unknowns $K_p, T_i$ and $T_d$. A possibility to carry
out the design at this point is to freely assign one of the unknowns
and to solve (\ref{eqaa}-\ref{eqbb}) for the other
two.\footnote{From (\ref{eqaa}) it is easily seen that $K_p$ cannot
be chosen arbitrarily. If we choose $T_i$, (\ref{eqaa}) gives
$K_p=M_g\,\cos \varphi_g$, and (\ref{eqbb}) leads to
\[
T_d=\dfrac{1+\omega_g\,T_i\,\tan \varphi_g}{T_i\,\omega_g^2}.
\]
However, in order to guarantee $K_p>0$ and $T_d>0$, the angle
$\varphi_g$ must be such that $\cos \varphi_g>0$ and $T_i$ must be
chosen to be smaller than $-1/(\omega_g\,\tan \varphi_g)$. These two
conditions can be simultaneously satisfied only when $\varphi_g \in
(-\pi/2, 0)$. If we choose $T_d$, we get $K_p=M_g\,\cos \varphi_g$
and
\[
T_i=\dfrac{1}{\omega_g^2\,T_d-\omega_g\,\tan \varphi_g},
\]
which implies that in order to ensure $T_i>0$ we must choose $T_d$
to be greater than $\tan \varphi_g / \omega_g$. Therefore, $T_d$ is
arbitrary when $\varphi_g \in (-\pi/2, 0)$, while when $\varphi_g
\in (0,\pi/2)$, we must choose $T_d>\tan \varphi_g / \omega_g$.}

Another possibility is to exploit the remaining degree of freedom so
as to satisfy some further time or frequency domain requirements.
Here, we consider two important situations: the first is the one
where the ratio $T_d/T_i$ is chosen to ensure that the zeros of the
PID controller are real; the second is the one where a gain margin
constraint is to be satisfied.

\subsubsection{Imposition of the ratio $T_d/T_i$}

The ratio $\sigma\stackrel{\text{\tiny def}}{=}T_d/T_i$ is an
important parameter. When $\sigma^{-1} \ge 4$, the zeros of the PID
controller are real, and they are complex conjugate when
$\sigma^{-1} < 4$. In the following theorem, necessary and
sufficient conditions are given for the solvability of
(\ref{inversion}) in the case of a standard PID controller when the
ratio $\sigma$ is given, \cite{Astrom-06}.
%Moreover, closed-form formulae are provided for the parameters of the PID controller to meet the specifications on $\textrm{PM}$, $\omega_g$ and $\sigma$ exactly.

\begin{theorem} {(\cite[Ch. 4, pp. 140--141]{Astrom-06}).}
\label{the1} Let $\sigma=T_d/T_i$ be assigned. Equation
(\ref{inversion}) with $\bar{C}(s)=C_{\rm PID}(s)$ admits solutions
in $K_p, T_i, T_d >0$ if and only if $\varphi_g \in ( -{\pi}/{2},
{\pi}/{2})$. If this condition is satisfied, the solution of
(\ref{eqpar}) is given by
 \bea
\phantom{PPPP} K_p & = & M_g\,\cos \varphi_g \label{eqKp}\\
T_i & = & \dfrac{\tan \varphi_g+ \sqrt{\tan^2 \varphi_g + 4\,\sigma}}{2\,\omega_g\,\sigma} \phantom{ploplo}\label{eqTi}\\
T_d & = & T_i\,\sigma  \label{eqTd}
 \eea
\end{theorem}
\proof {\bf (Only if)}. As already observed, equating real part to
real part and imaginary part to imaginary part in (\ref{eqpar})
results in (\ref{eqaa}) and (\ref{eqbb}). Since $K_p$ must be
positive, from (\ref{eqaa}) -- which can be written as
$K_p=M_g\,\cos \varphi_g$ -- we get that $\varphi_g$ must satisfy
$-\pi/2 < \varphi_g < \pi/2$. If this inequality is satisfied, it is
also easy to see that (\ref{eqbb}) always admits a positive
solution. In fact, (\ref{eqbb}) can be written as
 \bea
 \label{eqaabbb}
\omega_g^2\,\sigma\,T_i^2-\omega_g\,T_i\,\tan \varphi_g -1=0,
 \eea
in $T_i$, that always admits two real solutions, one positive and one negative. \\
{\bf (If)}. From (\ref{eqKp}), it follows that (\ref{eqaa}) is
satisfied. Moreover, since as aforementioned (\ref{eqbb}) can be
written as (\ref{eqaabbb}) and $\sqrt{\tan^2 \varphi +
4\,\sigma}>|\tan \varphi|$, the positive solution is given by
(\ref{eqTi}).
\endproof

\subsubsection{Imposition of the Gain Margin}
Another possibility in the solution of the control problem in the
case of unconstrained $K_i$ is to fix the gain margin to a certain
value $\textrm{GM}$. From $\textrm{arg}\,L(j \omega_p) = -\pi$ and
$\textrm{GM}  = {|L(j \omega_p)|}^{-1}$ we obtain $M_p=
{1}/({\phantom{\Big|}\textrm{GM}\,|\bar{G}(j
\omega_p)|\phantom{\Big|}})$ and
$\varphi_p=-\pi-\textrm{arg}\,\bar{G}(j \omega_p)$. Therefore, now
the parameters $K_p, T_i, T_d>0$ of the PID controller must be
determined so that (\ref{eqpar}) and
 \bea
\label{eqpar1}
M_p\,e^{\,j\,\varphi_p}=K_p\,\dfrac{1+j\,\omega_p\,T_i-\omega_p^2\,T_i\,T_d}{j\,\omega_p\,T_i}
\eea are simultaneously satisfied. By equating real and imaginary
part of (\ref{eqpar}) and (\ref{eqpar1}) we obtain (\ref{eqaa}),
(\ref{eqbb}) in addition to \bea
&& \omega_p\,M_p\,T_i\,\cos \varphi_p = \omega_p\,K_p\,T_i, \label{eqcc} \\
&& -M_p\,\omega_p\,T_i\,\sin \varphi_p = K_p-K_p\,\omega_p^2\,T_i\,T_d.\label{eqdd}
 \eea
From (\ref{eqaa}) and (\ref{eqcc}), we obtain
 \bea
\label{const1}
M_g\,\cos \varphi_g=M_p\,\cos \varphi_p
 \eea
in the unknown $\omega_p$. For the control problem to be solvable,
it is required that (\ref{const1}) admits at least one strictly
positive solution. For the given $G(s)$ the solution of
(\ref{const1}) can be found by solving a polynomial equation in
$\omega_p$, and therefore {\em all} its solutions can be determined
either in closed form when the system order is not too hight or
numerically with arbitrary precision. Indeed, as for Lead-Lag
networks, it is a simple exercise of trigonometry to see that if
$G(s)$ is a rational function in $s \in \mathbb{C}$, (\ref{const1})
is a polynomial equation in $\omega_p$. If the transfer function of
the process is given by the product of a rational function
$\hat{G}(s)$, and a delay $e^{-t_0\,s}$, i.e., if
$G(s)=\hat{G}(s)\,e^{-t_0\,s}$, equation (\ref{const1}) is not
polynomial in $\omega_p$ and it needs to be solved numerically.

\begin{theorem}
\label{then} Consider Problem \ref{pro3} with the additional
specification on the gain margin $\textrm{GM}$. Equations
(\ref{eqpar}) and (\ref{eqpar1}) admit solutions in $K_p, T_i,
T_d>0$ if and only if $\varphi_g \in (-\pi/2,\pi/2)$ and
(\ref{const1}) admits a positive solution $\omega_p$ such that
 \bea
\label{conditions}
\left\{ \begin{array}{ll}
\! \omega_p < \omega_g \\
\! \omega_g\,\tan \varphi_g> \omega_p\,\tan \varphi_p \\
\! \omega_p \tan \varphi_g >  \omega_g\,\tan \varphi_p
\end{array} \right.
\quad \! \!\!\! \textrm{or} \!\quad \left\{ \begin{array}{ll}
\! \omega_p > \omega_g \\
\! \omega_g\,\tan \varphi_g < \omega_p\,\tan \varphi_p \\
\! \omega_p \tan \varphi_g <  \omega_g\,\tan \varphi_p \end{array} \right.
\eea
If $\varphi_g \in (-\pi/2,\pi/2)$ and (\ref{conditions}) is satisfied, the problem admits solutions with
\bea
 \phantom{PPPP} K_p & = &  M_g\,\cos \varphi_g \label{par1} \\[2mm]
 T_i & = & \frac{\omega_g^2-\omega_p^2}{\omega_g\,\omega_p\,\left( \omega_p\,\tan \varphi_g-\omega_g\,\tan \varphi_p\right)} \phantom{pippopl}  \label{par2} \\[3mm]
 T_d & = & \frac{\omega_g\,\tan \varphi_g-\omega_p\,\tan \varphi_p}{\omega_g^2-\omega_p^2}  \label{par3}
 \eea
\end{theorem}
\proof {\bf (Only if).} As already seen, a necessary condition for
the problem to admit solutions is that $\omega_p$ is a solution of
(\ref{const1}). From (\ref{eqaa}) and (\ref{eqbb}), and from
(\ref{eqcc}) and (\ref{eqdd}), we obtain
 \bea
 %\label{eqccdd}
 -\omega_g\,T_i\,\tan \varphi_g=1-\omega_g^2\,T_i\,T_d, \label{eqaabb} \\
 -\omega_p\,T_i\,\tan \varphi_p=1-\omega_p^2\,T_i\,T_d. \label{eqccdd}
 \eea
By solving (\ref{eqaabb}) and (\ref{eqccdd}) in $T_i$ and $T_d$, we
obtain (\ref{par1}-\ref{par3}).  For $K_p$ to be positive, it is
necessary that $\varphi_g \in (-\pi/2,\pi/2)$. Moreover, the time
constants $T_i$ and $T_d$ are positive if $\omega_g$ and $\omega_p$
satisfy (\ref{conditions}).
\\
 {\bf (If).} It is a matter of straightforward substitution of (\ref{par1}-\ref{par3}) into (\ref{eqaa}), (\ref{eqbb}), (\ref{eqcc}) and (\ref{eqdd}).
 \endproof

\begin{figure}[htb]
\centering
\psfrag{R}[]{\small $R(s)$}
\psfrag{E}[]{\small $\,$}
\psfrag{G}[]{\small $Y(s)$}
\psfrag{I}[]{\tiny $+$}
\psfrag{J}[]{\tiny $-$}
\psfrag{U}[]{\small $\,$}
\psfrag{E}[]{\small $E(s)$}
\psfrag{H}[]{\small $\,H(s)$}
\psfrag{B}[]{\small $\ssp \,1+T_i\,s+T_i\,T_d\,s^2\ssp$}
\psfrag{A}[]{\small $\ssp \,\, \dfrac{K_p}{T_i\,s}\,G(s)  \ssp$}
\includegraphics[width=3.3in]{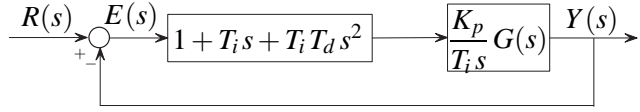}
\caption{Modified feedback structure with unity DC gain controller.}
\label{fig4}
\end{figure}

Now we consider the case in which the steady-state requirements lead
to constraints on $K_i=K_p/T_i$.
Hence, now the integration constant $K_i$ is determined via the
imposition of the steady-state requirements; for example, for type-0
(resp. type-1) plants, $K_i$ is computed via the imposition of the
velocity error (resp. acceleration error).
\\[2mm]
As such, the factor $K_i/s$ can be separated from $\bar{C}_{\rm
PID}(s)=1+T_i\,s+T_i\,T_d\,s^2$, and viewed as part of the plant. In
this way, the part of the controller to be designed is $\bar{C}_{\rm
PID}(s)$, and the feedback scheme reduces to that of Figure
\ref{fig4}.
%%%%%%%%%%%%%%%%%%%%%%%%%%%%%%%%%%%%
%%%%%%%   FIGURE   %%%%%%%%%%%%%%%%%
%%%%%%%%%%%%%%%%%%%%%%%%%%%%%%%%%%%%
%
%Let
%\bea
%\label{defL}
%L(s)=\tilde{C}_{\rm PID}(s)\,\Big(\frac{K_p}{T_i\,s}\,G(s)\Big)
%\eea
 %denote the loop gain transfer function.
 Let $\tilde{G}(s)\stackrel{\text{\tiny def}}{=}\frac{K_p}{T_i\,s}\,G(s)$, so that the loop gain transfer function can be written as $L(s)=\bar{C}_{\rm PID}(s)\,\tilde{G}(s)$. Write $\tilde{G}(j \omega)$ and $\bar{C}_{\rm PID}(j \omega)$ in polar form
 as
 \beann
\tilde{G}(j \omega)= |\tilde{G}(j \omega)|\,e^{\,j\,\textrm{arg}\,\tilde{G}(j \omega)}, \qquad  \bar{C}_{\rm PID}(j \omega)=M(\omega)\,e^{\,j\,\varphi(\omega)},
 \eeann
so that the loop gain frequency response can be written as $L(j
\omega)= |\tilde{G}(j
\omega)|\,M(\omega)\,e^{\,j\,\left(\textrm{arg}\,\tilde{G}(j
\omega)+\varphi(\omega)\right)}$. If the crossover frequency
$\omega_g$ and the phase margin $\textrm{PM}$ of the loop gain
transfer function $L(s)$ are assigned, the equations $|L(j
\omega_g)| = 1$ and $\textrm{PM}  = \pi+\textrm{arg}\,L(j \omega_g)$
must be satisfied, and as already observed these can be written as
$M_g= {1}/{\phantom{\Big|}|\tilde{G}(j \omega_g)|\phantom{\Big|}}$
and $\varphi_g=\textrm{PM}-\pi-\textrm{arg}\,\tilde{G}(j \omega_g)$.
Alternatively, $M_g$ and $\varphi_g$ can be computed as functions of
the frequency response of $G(s)$ at $\omega=\omega_g$:
 \bea
M_g & = & \left| \frac{K_p}{T_i\,j\,\omega_g}\,G(j \omega_g) \right|^{-1} = \dfrac{\omega_g}{K_i \,\left| G(j \omega_g) \right|} \label{defM} \\
\varphi_g  & =  & \textrm{PM}-\pi-\textrm{arg}\,\left[ \frac{K_p}{T_i\,j\,\omega_g}\,G(j \omega_g) \right]  \nonumber\\
  %& =  &
    &  =  & \textrm{PM}-\frac{\pi}{2}- \textrm{arg}\, G(j \omega_g), \label{defphi}
 \eea
 since $K_p,T_i > 0$.
%, respectively. We get
%\begin{enumerate}
%\item $|\tilde{G}(j \omega_g)|\,M(\omega_g) = 1 \quad \implies \quad M(\omega_g)= \dfrac{1}{\phantom{\Big|}|\tilde{G}(j \omega_g)|\phantom{\Big|}}$,\\[-5mm]
%\item $\textrm{PM}  = \pi+\textrm{arg}\,\tilde{G}(j \omega_g)+\varphi(\omega_g) \quad \implies \quad \varphi(\omega_g)=\textrm{PM}-\pi-\textrm{arg}\,\tilde{G}(j \omega_g)$.
%\end{enumerate}
In order to find the parameters of the controller such that {\em
i-a)} and {\em ii-a)} are met, equation \bea \label{eqpid} M_g\,e^{j
\,\varphi_g}= 1+j\,\omega_g\,T_i-T_i\,T_d\,\omega_g^2 \eea must be
solved in $T_i>0$ and $T_d>0$. The closed-form solution to this
problem is given in the following theorem.
\begin{theorem}
\label{the2}
Equation (\ref{eqpid}) admits solutions in $T_i>0$ and $T_d>0$ if and only if
\bea
\label{conditionspid}
0<\varphi_g<\pi \qquad \textrm{and}\quad M_g\,\cos \varphi_g < 1.
\eea
If (\ref{conditionspid}) are satisfied, the solution of (\ref{eqpid}) is given by\\[-2mm]
\bea
K_p & = & K_i\,\dfrac{1}{\omega_g}\,M_g \sin \varphi_g, \phantom{ploplo} \\
T_i & = & \dfrac{1}{\omega_g}\,M_g \sin \varphi_g, \label{parpida}\\
 T_d & = & \dfrac{1-M_g\,\cos \varphi_g}{\omega_g\,M_g\,\sin \varphi_g}. \label{parpidb}
\eea
\end{theorem}
\ \\[-2mm]
%Now we can compute $K_p$ with the value of $T_i$ thus found, from the ratio $K_i=K_p/T_i$.
%Notice also that we can express the solvability conditions and the parameters of the controller in terms of the modulus and argument of $G(j \omega)$ at $\omega=\omega_g$. We can express these parameters as
%\beann
%T_i & = & -\frac{1}{K_i\,|G(j \omega_g)|}\,\cos \Big( \textrm{PM}-\textrm{arg}\,{G}(j \omega_g) \Big),\\
%T_d & = & \frac{{K_i}|G(j \omega_g)|- \omega_g\,\sin \Big( \textrm{PM}-\textrm{arg}\,{G}(j \omega_g) \Big)}{-\omega_g^2\,\cos \Big( %\textrm{PM}-\textrm{arg}\,{G}(j \omega_g) \Big)}.
%\eeann
The two conditions (\ref{conditionspid}) can be alternatively written as
\beann
\left. \begin{array}{ccc}
\varphi_g \in \left( \arccos \dfrac{1}{M_g}, \, \pi \right) & \textrm{if}\; M_g > 1, \\[1mm]
\varphi_g \in ( 0, \, \pi ) & \textrm{if}\; M_g < 1.
\end{array} \right.
\eeann In fact, when $\varphi_g \in (0,\pi/2)$, condition $\cos
\varphi_g < 1/M_g$ is always satisfied when $M_g<1$, and is
satisfied when $\varphi_g > \arccos (1/M_g)$ when $M_g>1$. When
$\varphi_g \in (\pi/2,\pi)$, the condition $\cos \varphi_g < 1/M_g$
is always satisfied since $\cos \varphi_g < 0$ and $(1/M_g)>0$.

\subsection{Design examples using PID controllers}
\label{ex1}
Consider the unity feedback control scheme in Figure \ref{fig3}.

{\bf Question 1}. Design a compensator that meets the following specifications:

\begin{itemize}
\item zero velocity error;
\item phase margin equal to $45^\circ$;
\item gain crossover frequency equal to $3$ rad/sec.
\end{itemize}

\begin{figure}[tbp]
  \centering
  \psfrag{Im}[c][c][0.7 ]{$Im$}
  \psfrag{Re}[c][c][0.7 ]{$Re$}
  \psfrag{G}[tl][tl][1.0]{\hspace{1mm}$G(j\omega)$}
 \psfrag{L}[tl][tl][1.0]{$L(j\omega)$}
 \psfrag{wp}[tl][tl][0.5]{\hspace{-3mm}$3$}
 \psfrag{wp2}[tl][tl][0.5]{\hspace{0mm}$3$}
 \psfrag{A}[tl][tl][0.8]{\hspace{1mm}$A$}
 \psfrag{B}[tl][tl][0.8]{\hspace{2mm}$B$}
  \includegraphics[width=7.3cm]{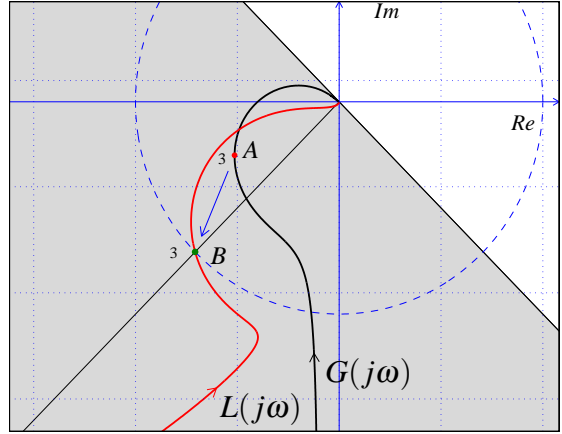}
\caption{Design of PID compensator on the Nyquist plane to meet the
specifications of Question $1$.}\label{PID_Question_1}
\end{figure}

The steady-state specification is automatically satisfied by using a
PID controller or a PI controller. Let us consider the case of a PID
controller. The extra freedom in this case can be used to select the
ratio $T_i/T_d$. Let us choose for example $T_i/T_d=8$, so that
$\sigma=1/8$ guarantees that the zeros of the PID controller are
real. Then, we compute $M_g$ and $\varphi_g$: \beann
M_g \ns&\ns = \ns&\ns \dfrac{1}{|{G}(3\,j)|} = 3 \sqrt{\dfrac{37}{109}}\simeq 1.7479, \\
\varphi_g \ns&\ns = \ns&\ns \textrm{PM}-(\pi+\textrm{arc} \bar{G}(3\,j)) \\
\ns&\ns = \ns&\ns \frac{7}{4}\pi-\arctan \frac{3}{10} + \arctan 6 \simeq 18.84^\circ.
\eeann
Since $\varphi_g\in (-\pi/2,\pi/2)$, the problem admits a solution with a PID controller. Using (\ref{eqKp}-\ref{eqTd}) we find
$K_p=1.6542$, $T_i=1.5017$ sec and $T_d=0.1877$ sec. This choice guarantees that the controller has real zeros, which in this case are
$-4.5471$ and $-0.7802$.

Let us attempt to solve the same problem with a PI controller. To this end, we compute
\[
\varphi_g=\textrm{PM}-\frac{\pi}{2}-\textrm{arg} G(j \omega_g) \simeq 108.8384^\circ,
\]
so that a PI controller solving the problem does not exist.

The graphical representation of the problem solution with PID
regulator is shown in Fig.~\ref{PID_Question_1}. The designed PID
brings the point $A=G(j\omega_g)$ to $B=e^{j\,(\textrm{PM}+\pi)}$.
The gray area corresponds to the set of admissible points that can
be brought to $B$ by a PID controller.

{\bf Question 2}. Design a compensator that meets the following specifications:

\begin{itemize}
\item zero velocity error and acceleration error not greater than $0.2$;
\item phase margin equal to $45^\circ$;
\item gain crossover frequency equal to $3$ rad/sec.
\end{itemize}

\begin{figure}[tbp]
  \centering
  \psfrag{Im}[c][c][0.7 ]{$Im$}
  \psfrag{Re}[c][c][0.7 ]{$Re$}
  \psfrag{G}[tl][tl][1.0]{\hspace{1mm}$\widetilde{G}(j\omega)$}
 \psfrag{L}[tl][tl][1.0]{$L(j\omega)$}
 \psfrag{wp}[tl][tl][0.5]{\hspace{3mm}$3$}
 \psfrag{wp2}[tl][tl][0.5]{\hspace{0mm}$3$}
 \psfrag{A}[tr][tr][0.8]{\hspace{-4mm}$A$}
 \psfrag{B}[tl][tl][0.8]{\hspace{2mm}$B$}
  \includegraphics[width=7.3cm]{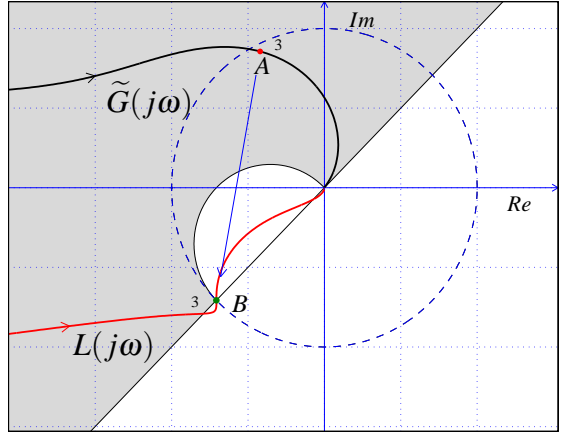}
\caption{Design of PID controller on the Nyquist plane to meet the
specifications of Question $2$.}\label{PID_Question_2}
\end{figure}
The correct compensator structure to be employed in this case is the PID controller.
 As already observed, the steady-state requirement in this case imposes the ratio $K_i=K_p/T_i$. In particular, in this case we need
 $K_i \ge 5$. Let us choose $K_i=5$. Hence,
 \beann
 M_g \ns&\ns = \ns&\ns \dfrac{\omega_g}{K_i\,|G(j \omega_g)|} \simeq 1.0487, \\
 \varphi_g \ns&\ns = \ns&\ns \textrm{PM}-\dfrac{\pi}{2}-\textrm{arg} G(j \omega_g) \simeq 108.8384^\circ.
 \eeann
 The conditions $0 < \varphi_g < \pi$ and $M_g\,\cos \varphi_g<1$ are both satisfied, so that the problem admits solutions. The parameters of the controller in this case are
 \beann
 T_i \ns&\ns = \ns&\ns \dfrac{M_g \sin \varphi_g}{\omega_g} \simeq  0.3308 \,\textrm{sec} \\
 T_d \ns&\ns = \ns&\ns \dfrac{1-M_g\,\cos \varphi_g}{\omega_g\,M_g\,\sin \varphi_g} \simeq 0.4496\,\textrm{sec} \\
 K_p \ns&\ns = \ns&\ns K_i\,T_i \simeq 1.6542.
 \eeann
 As such, the PID controller
 \beann
 C_{\rm PID}= 1.6542\left( 1+ \dfrac{1}{0.3308\,s}+0.4496\,s \right)
 \eeann
 solves the control problem. However, since in this case $T_i<4\,T_d$, the zeros of the compensator are complex conjugate and equal to $-1.1122 \pm 2.3423\,j$.
The Nyquist plot of $\tilde{G}(j \omega)$ is shown in
Fig.~\ref{PID_Question_2}. It can be shown that $\tilde{G}(j
\omega)$ is a rotation and an amplification of $G(j\omega)$ and the
area of points that can be brought to $B$ by the controller is shown
in gray.

\section{Conclusion}
In this paper we presented a design method for all types of standard
compensators that are ubiquitously addressed in Control subjects,
and which represent the very vast majority of compensators used in
industry. This method, based on the so-called {\em Inversion
Formulae}, enables the synthesis to be carried out precisely and
just with the aid of a pen and a piece of paper. This represents the
most remarkable value and potential of this method in Control
education. In fact, these techniques -- that have been employed in
several Universities in Italy for several years -- do not rely on
iterative procedures to be performed on Bode or Nyquist plots, and
appear therefore to be very suitable for numerical exercises that
can test students' skills in every single aspect of the compensator
design process. In this paper we tried to focus our attention on the
most important educational aspects of this technique, emphasizing
the links that can be established with the classic diagrams of the
frequency response, because we firmly believe that this is a key
aspect for a deep understanding of Control synthesis techniques.
However, the value of this method lies also in the fact that can be
easily adapted to different design scenarios:
\begin{itemize}
\item Here for the sake of brevity we restricted our attention to standard PID controllers. However, these techniques are easily adapted to PI and PD controllers, as well as to PID controllers with an additional pole introduced in the derivative action for physical implementability, \cite{Ntogramatzidis-F-11};
\item The approach based on the Inversion Formulae can easily be adapted to the discrete-time case as shown in \cite{Zanasi-M-09};
\item Even if in the examples proposed in this paper the transfer function was rational, it is very easy to see that this method can be readily applied to systems with finite delays as well.
\end{itemize}
The relevance of this method for written exercises has been
demonstrated in this paper with a number of Question examples that
are extremely difficult to tackle with the standard approaches, and
that shed some light on some aspects of the control design that
would otherwise remain neglected. In particular, we have proposed
several different design exercises aimed at designing the parameters
of the compensator (even in closed-form as a further evidence of the
simplicity of the method) in the presence of standard specifications
on the steady-state performance and stability margins and crossover
frequencies. This method can even provide an {\em a priori} answer
to the question if the desired stability margin guarantees
asymptotic stability of the closed-loop, by combining the Inversion
Formulae with the Routh criterion, while often it is with an {\em a
posteriori} check that in Control courses students are encouraged to
ensure a positive margin indeed leads to asymptotic stability.

%\appendices
\section*{Appendix A: Relationship between the two Lead-Lag network transfer functions}
The relationship between $C_{\rm LL}(s)$ and $C_{\rm LL}^\prime(s)$
can be proved by writing $C_{\rm LL}(s)$ as
 \beann
C_{\rm LL}(s)=
\dfrac{s^2+\dfrac{\tau_1+\tau_2}{\tau_1\,\tau_2}\,s+\dfrac{1}{\tau_1\,\tau_2}}{s^2+\dfrac{\alpha^2\,\tau_1+\tau_2}{\alpha\,\tau_1\,\tau_2}\,s+\dfrac{1}{\tau_1\,\tau_2}}
 \eeann
and by comparing this expression with $C_{\rm LL}^\prime(s)$.
Expressions (\ref{legame}) immediately follow. This shows that
$C_{\rm LL}(s)$ can always be written as $C_{\rm LL}^\prime(s)$. To
prove the opposite implication, we solve (\ref{legame}) in $\alpha$,
$\tau_1$ and $\tau_2$. Solving the third of (\ref{legame}) gives
 \bea
\tau_1=1/(\tau_2\,\omega_n^2). \label{t1}
 \eea
 Plugging this into the first of (\ref{legame}) gives
\[
\zeta_1=\dfrac{1+\tau_2^2\,\omega_n^2}{2\,\tau_2\,\omega_n},
\]
which leads to the equation in $\tau_2$:
\[
\tau_2^2\,\omega_n^2-2\,\zeta_1\,\tau_2\,\omega_n+1=0,
\]
whose solutions are $\tau_2^{\pm}=\hat{\zeta_1}^{\pm}/\omega_n$. To
find the first set of solutions, let us first consider
$\tau_2=\tau_2^+$, and we plug it into (\ref{t1}) to get
$\tau_1=\hat{\zeta}_1^-/\omega_n$. We plug these expressions in the
second of (\ref{legame}) and we get
 \beann
\alpha^2\zeta_1^--2\,\zeta_2\,\alpha+\zeta_1^+=0
 \eeann
which gives $\alpha=\hat{\zeta}_2^{\pm}/\hat{\zeta}_1^-$. As such,
the first solution yields $\tau_2=\hat{\zeta}_1^+/\omega_n$,
$\tau_1=\hat{\zeta}_1^-/\omega_n$ and
$\alpha=\hat{\zeta}_2^{\pm}/\hat{\zeta}_1^-$. These two solutions
both lead to $\tau_1>0$ and $\tau_2>0$. However, in the case
$\zeta_2>\zeta_1>1$, the only solution that gives $\alpha \in (0,1)$
is $\tau_2=\hat{\zeta}_1^+/\omega_n$,
$\tau_1=\hat{\zeta}_1^-/\omega_n$ and
$\alpha=\hat{\zeta}_2^{-}/\hat{\zeta}_1^-$, as one can see with
simple irrational inequalities. When $\zeta_1>\zeta_2>1$, both
solutions lead to $\alpha \in (0,1)$, and are therefore both
feasible. The second set of solutions is found by picking
$\tau_2=\tau_2^-$. By following the same steps, we get
$\tau_2=\hat{\zeta}_1^-/\omega_n$, $\tau_1=\hat{\zeta}_1^+/\omega_n$
and $\alpha=\hat{\zeta}_2^{\pm}/\hat{\zeta}_1^+$. When
$\zeta_1>\zeta_2>1$, $\alpha$ is always greater than $1$. When
$\zeta_2>\zeta_1>1$, the only solution that yields $\alpha \in
(0,1)$ is $\tau_2=\hat{\zeta}_1^+/\omega_n$,
$\tau_1=\hat{\zeta}_1^-/\omega_n$ and
$\alpha=\hat{\zeta}_2^{-}/\hat{\zeta}_1^-$. \endproof

%Appendix one text goes here.

%\section{}
%Appendix two text goes here.

%\section*{Acknowledgment}

% Can use something like this to put references on a page
% by themselves when using endfloat and the captionsoff option.
%\ifCLASSOPTIONcaptionsoff
%  \newpage
%\fi

%

\end{document}
\end{document}